\documentclass[useAMs,a4paper]{mn2e}
\usepackage{savesym}
\usepackage{graphicx}
\expandafter\let\csname equation*\endcsname\relax
  \expandafter\let\csname endequation*\endcsname\relax 
\usepackage{subfig}
\usepackage{amsmath}
\DeclareMathOperator\arccosh{arccosh}

\usepackage{amssymb}
\usepackage{verbatim}
\usepackage[yyyymmdd,hhmmss]{datetime}
\usepackage{array}
\usepackage{times}
\usepackage[total={17.8cm,24.0cm},centering]{geometry} 
\usepackage{color}

\newcommand{\beq}{\begin{equation}}
\newcommand{\eeq}{\end{equation}}

\newcommand\W {{W^r_\phi}}

\title[ Green's functions for relativistic discs]{ Asymptotic  Green's function solutions of the general relativistic  thin  disc equations   }
\author [Andrew Mummery]{Andrew Mummery\thanks{E-mail:
andrew.mummery@physics.ox.ac.uk}
\\
Oxford Astrophysics, Denys Wilkinson Building, Keble Road, Oxford, OX1 3RH, United Kingdom}
\begin{document}

\date{}

\pagerange{\pageref{firstpage}--\pageref{lastpage}} \pubyear{2022}

\maketitle

\label{firstpage}

\begin{abstract} 
The leading order Green's function solutions of the  general relativistic thin disc equations are computed, using a pseudo-Newtonian potential and asymptotic Laplace mode matching techniques.    This solution, valid for a vanishing ISCO stress, is constructed by ensuring that it reproduces the leading order asymptotic behaviour of the near-ISCO, Newtonian, and global WKB limits.  Despite the simplifications used in constructing this solution, it is typically accurate, for all values of the Kerr spin parameter $a$ and at all radii,  to less than a percent of the full numerically calculated solutions of the general relativistic disc equations. These solutions will be of use in studying time-dependent accretion discs surrounding Kerr black holes.  
\end{abstract}

\noindent

\section{Introduction}
In the classical theory of thin accretion discs, the constraints of angular momentum and mass conservation may be combined into a single evolutionary equation for the disc's surface density, a  result first  discussed at length by Lynden-Bell \& Pringle (1974). Quite generally the solutions of this evolutionary equation show that the bulk of the matter in the accretion disc drifts inward, while the angular momentum of the disc is transported outward, sustained by a vanishingly small mass fraction of the disc.  

One of the key results provided by Lynden-Bell \& Pringle was a set of Green's function solutions of the Newtonian evolutionary equation, which exist provided that the disc's turbulent stress tensor $\W$ follows a power-law with disc radius $r$. A Green's function is an extremely powerful analytical tool, as it describes the evolution of the disc density defined by a $\delta$-function initial condition.  The solution of the general problem defined by an arbitrary initial condition is then simply given by an integral of this initial condition with the Green's function solution.  
Indeed, the Green's functions of Lynden-Bell  \& Pringle have been widely used throughout the astrophysical literature, including in instances where they are not strictly valid. In particular, these Newtonian Green's function solutions will not provide an accurate description of the inner regions of those discs evolving around black holes. Black hole discs must terminate at a finite radius given by the innermost stable circular orbit (ISCO), and will  be  modified by the changes to the orbital velocities of the disc fluid in the strong gravity regions of the black hole's spacetime.  

The  disc equation which describes the evolution of the disc surface density in full general relativity was first written down by Eardley \& Lightman (1975), before being rediscovered in slightly different coordinates by Balbus  (2017). Balbus (2017) presented a set of formal  Wentzel--Kramers--Brillouin (WKB) Laplace mode solutions of this equation, including a modal solution which is universally valid (i.e., a solution which reproduces the leading order solutions of the disc equations in each of the small radius, large radius and WKB limits). From this modal solution, a formal asymptotic Green's function can be constructed. 

There is, however, more than one modal solution which can be constructed to reproduce the required leading order asymptotic  behaviour. The particular Balbus (2017) solution has a late-time leading order behaviour of $\sim 1/t^3$, which is  steeper than that found for the numerical solutions of the relativistic disc equations (Balbus \& Mummery 2018).     In addition, the formal solutions of Balbus (2017) contain integrals which do not have closed form solutions for the Kerr metric.  It is therefore of interest to seek a set of simpler Green's function solutions,  which accurately capture the properties of the full numerical solutions of the relativistic disc equations, but which can be described in a closed form.  

In this paper we derive a set of such Green's function solutions.   The techniques employed are similar in style to those of Balbus (2017), but differ in the details.  We simplify the relativistic disc equations, leaving a problem with solutions which can be written in a closed form but which still capture the leading order relativistic physics relevant for accretion.  The principal simplification is the use of a pseudo-Newtonian potential which reproduces the Kerr spacetime's ISCO location, while producing  technically simpler fluid orbital motion. A global modal solution is then found for this pseudo-potential's  disc evolution equation which reproduces the  correct asymptotic behaviour  in each of the near-ISCO, Newtonian and global WKB limits.  This modal solution can then be used to construct an asymptotic Green's function.    Despite the simplifications used in constructing this solution it is typically accurate, for all values of the Kerr spin parameter $a$ and at all radii,  to less than a percent of the full numerically calculated solutions of the general relativistic disc equations. 

We then analyse the properties of this asymptotic Green's function. We show that the mass accretion rate across the ISCO has a particularly simple closed form, which can be used to understand the black hole spin dependence of the inner-disc accretion rate.  The mass accretion rate across the ISCO is, for fixed disc mass, {\it lower} at peak for more rapidly rotating Kerr black holes. However, the ISCO accretion rate remains higher as a fraction of its peak value for much {\it longer} for these spinning black holes.  When the differing mass-to-light efficiencies of different Kerr spacetimes are taken into account, the rest-frame luminosities of the more rapidly spinning black holes are largest.  

The layout of this paper is as follows: in section \ref{sec2} we motivate and set up the problem to be solved.    In section \ref{sec3} the asymptotic Green's function solution of the disc equations is derived. These solutions are then compared to the numerical solutions of the full relativistic disc equations  in section \ref{num}, where close agreement between the two is found. In section \ref{sec5} the properties of the Green's function of the mass accretion rate are analysed, we then conclude in section \ref{sec6}.  Technical results pertinent to the analysis are presented in four appendices. 

\section{Preliminary analysis}\label{sec2}
In this section we both motivate and set up the problem to be solved in the remainder of this paper. We begin by recapping the properties of the general relativistic thin disc evolution equation,  highlighting its prohibitive algebraic complexity which prevents  exact Green's function solutions from being found. Turning to a simpler problem, we introduce a pseudo-Newtonian potential which captures the leading order general relativistic effects relevant for studying thin accretion discs. With a suitable parameterisation of the turbulent stress tensor $\W$, the governing Laplace mode equation of this model is derived, which is then solved in section \ref{sec3}.

\subsection{The general relativistic disc equation}
The coordinates used to describe the relativistic thin disc equation are the cylindrical Boyer-Lindquist representation of the Kerr metric: $r$ (radial), $\phi$ (azimuthal), and $z$ (vertical).   The governing equation describes the evolution of the azimuthally-averaged, height-integrated disc surface density $\Sigma (r, t)$.   The contravariant four velocity of the disc fluid is $U^\mu$; the covariant counterpart is $U_\mu$.  The specific angular momentum corresponds to $U_\phi$, a covariant quantity.     There is an anomalous stress tensor present, $\W$, due to low level disk turbulence, which is a measure of the correlation between the fluctuations in $U^r$ and $U_\phi$ (Eardley \& Lightman 1975, Balbus 2017).  This is, as the notation suggests, a mixed tensor.    $\W$ serves both to transport angular momentum as well as to extract the free-energy of the disc shear, which is then thermalised and radiated from the disc surface, both assumed to be local processes.    

Under these assumptions the governing disc equation can be expressed in the following compact form 
\beq\label{27q}
{\partial \zeta\over \partial t} =  { \W\over (U^0)^2}{\partial\ \over \partial r} \left({U^0\over U'_\phi} \left[   {\partial \zeta \over \partial r}\right] \right).
\eeq
here the primed notation $'$ denotes an ordinary derivative with respect to $r$, and 
\beq
\zeta \equiv {r \Sigma \W \over U^0} .
\eeq 

We see that one only needs expressions for two of the orbital components of the disc's flow to solve this equation. These are the time dilation factor 
 \begin{equation}
U^0 = \frac{1+a\sqrt{{r_g}/{r^3}}}{\left({1 - {3r_g}/{r} + 2a\sqrt{{r_g}/{r^3} } }\right)^{1/2}} ,
\end{equation}
and the circular orbit angular momentum gradient 
 \beq\label{rel_ang_mom}
 U_\phi ' = \frac{\sqrt{GM} \left( a\sqrt{r_g} + r^{{3}/{2}} \right) \left( r^2 - 6r_g r - 3a^2 + 8a\sqrt{r_g r}\right)}{2r^4 \left(  1 - {3r_g}/{r} + 2a\sqrt{{r_g}/{r^3}}   \right)^{{3}/{2}}}.
 \eeq
 In these expressions, $a$ is the black hole's angular momentum parameter (having dimensions of length), $M $ is the black hole's mass, $r_g = GM/c^2$ the gravitational radius, and $G$ and $c$ are Newton's constant and the speed of light respectively. 
 
 Clearly, the relativistic disc equation (\ref{27q}) is extremely algebraically complex. Formal WKB Laplace mode solutions of eq. (\ref{27q}) were written down by Balbus (2017), but contain integrals of the form 
 \beq
 X = \int \left({U^0 U_\phi ' \over \W } \right)^{1/2} \, {\rm d}r, 
 \eeq
 which do not,  for the general Kerr metric, have closed form solutions. The question addressed in this paper is whether closed form solutions of a closely related problem can be found which accurately reproduce the behaviour of the numerical solutions of equation (\ref{27q}), while being much simpler to implement. 
 
 It is important therefore to understand what exactly the key properties of equation (\ref{27q}) are before we move forward.  The function $U^0$ reduces to unity in the Newtonian limit, and over the entire domain of interest in the Kerr geometry  is  smooth, non-vanishing, and bounded.   The physical content of equation (\ref{27q}) may thus be retained by ignoring this function, in effect setting it equal to unity.   By contrast, $U'_\phi$ vanishes at the ISCO, introducing an apparent singularity into the equation, and must be handled with more care.  The ISCO is the radial location (denoted $r_I$) where the factor 
 \beq
  r_I^2 - 6r_g r_I - 3a^2 + 8a\sqrt{r_g r_I} = 0, 
 \eeq
 and the angular momentum gradient transitions from positive to negative.  At this location, for all values of the Kerr  spin parameter, the angular momentum gradient locally behaves like 
 \beq
 U_\phi ' \sim A (r - r_I) ,
 \eeq
 for constant $A\geq 0$.  It is this behaviour, where the angular momentum gradient vanishes linearly at a special location, which is the gross leading order relativistic correction to a Newtonian  theory.   
 
Full general relativity is not required to produce this property of a linearly vanishing angular momentum gradient. Instead, simpler so-called `pseudo' Newtonian potentials can be constructed to reproduce this property in Newton's gravity, which we now discuss.   
 
\subsection{A pseudo-Newtonian potential}
The use of pseudo-Newtonian potentials to model some effects of general relativistic gravitational fields was initiated by Paczy{\'n}ski \& Wiita (1980), who used 
\beq
\Phi(r) = -{GM\over r- R_S}, \quad R_S \equiv {2GM \over c^2 }, 
\eeq  
 to numerically calculate models of thick disks. This potential  does not satisfy  Laplace's equation $\nabla^2 \Phi \neq 0$ (hence `pseudo' Newtonian), but is spherically symmetric, and quite useful in reproducing key features of orbital motion around a Schwarzschild black hole, such as the innermost stable orbit. This potential in effect captures the leading order  Schwarzschild effects most relevant to the study of accretion flows.

The Paczy{\'n}ski--Wiita potential is not the only potential function which reproduces these key relativistic effects, and is not necessarily the easiest to work with analytically. Consider the different potential function (which is a Kerr-generalised form of the potential first written down for Schwarzschild black holes by  Klu{\'z}niak \& Lee 2002)
\beq\label{PDef}
\Phi(r) = {GM \over r_I} \left[1 - \exp\left({r_I \over r}\right)\right] .
\eeq
For $r \gg r_I$ this potential reduces to the Newtonian point mass potential $\Phi(r) = -GM/r$. The Newtonian angular momentum of a circular orbit ($J$) is related to the gravitational potential via 
\beq
J = \left({r^3 {{\rm d} \Phi \over {\rm d} r}}\right)^{1/2} = \sqrt{GMr} \, \exp\left({r_I \over 2r}\right) .
\eeq
Differentiating this angular momentum with respect to the circular orbit radius $r$ leaves 
\beq\label{Jprime}
{{\rm d} J \over {\rm d} r} = {1\over 2} \sqrt{GM \over r} \exp\left({r_I \over 2 r}\right) \left[ 1 - {r_I \over r}\right] = {J \over 2 r} \left[ 1 - {r_I \over r}\right]. 
\eeq
We see that the circular orbit angular momentum gradient for the potential defined in (\ref{PDef}) goes to zero at $r = r_I$, and is negative for $r < r_I$. Circular orbits with a negative angular momentum gradient are unstable via the Rayleigh criterion, and so the potential defined in (\ref{PDef}) has an `ISCO' at the point $r = r_I$, and  can therefore be used to mimic important general relativistic effects, while being analytically much simpler to work with. 
 At large radii, the Newtonian angular momentum gradient $2 J' = \sqrt{GM/r}$ is recovered. 


The Newtonian disc equation which results from the pseudo-potential (\ref{PDef}) is the following
\beq\label{gov_pde}
{\partial y \over \partial t} = {2\W \over \sqrt{GM}} {\partial \over \partial r} \left[ {r^{1/2} \exp\left(-{r_I / 2r}\right) \over 1 - r_I/r } {\partial y \over \partial r} \right] ,
\eeq
where 
\beq
y \equiv r\Sigma \W. 
\eeq
This equation will form the basis of our analytical model.  

\subsection{Laplace modes}
We will seek Green's function solutions of equation (\ref{gov_pde}) by solving for its Laplace modes.  A Laplace mode is defined with a particularly simple time-dependence $y(x,t) = y(x,s) \, e^{-st}, s \geq 0$. Meaning that our governing modal equation has the form 
\beq
- s y = {2\W \over \sqrt{GM}} {\partial \over \partial r} \left[ {r^{1/2} \exp\left(-{r_I / 2r}\right) \over 1 - r_I/r } {\partial y \over \partial r} \right] . 
\eeq
The general time-dependent solution is then found by super-imposing modes with a weighting function 
\beq\label{laplacedef}
y(x,t ) = \int_0^\infty y(x,s) F(s) e^{-st} \, {\rm d}s,
\eeq
where $F(s)$ is set entirely by the initial conditions. For our purposes this initial condition will be a delta function $\delta(x-x_0)$. 

\subsection{Turbulent stress modelling} 
Essentially all analytic and semi-analytic treatments of accretion discs rely on the use of $\W$, the so-called `anomalous stress tensor' to account for the enhanced angular momentum transport and thermal energy dissipation associated with the accretion process. First invoked by Shakura and Sunyaev (1973), a measured turbulent stress tensor may now be extracted from numerical simulations of discs, as a consequence of the magnetorotational instability (Balbus and Hawley 1991). Even with explicit numerical studies however, turbulence remains at best poorly understood. There is no consensus fundamental theory that allows one, for example, to express this stress tensor in terms of background mean disc properties. 

In general the turbulent stress will depend on the local disc properties (temperature, pressure, density, etc.).  However, Newtonian Green's function solutions only exist when the stress is parameterised  by a power-law of disc radius $\W \propto r^\mu$, and this is therefore the most common parameterisation used for analytical studies of time-dependent thin discs. 

For the ease of solving the asymptotic disc equations it will in fact be of use to define the turbulent stress by 
\beq
\W = w \left({2r \over r_I}\right)^\mu \exp\left({r_I \over 2 r}\right) ,
\eeq
where $w$ is a constant which carries the dimensions of $\W$. The factor of $\exp(r_I/2r)$ in this  parameterisation is in effect a compensating `trick'  to cancel  factors of $\exp(r_I/2r)$ introduced by the pseudo-potential (eq.  \ref{PDef}). This `trick' results in maximum deviations from the traditionally used parameterisation of $\sqrt{e} \simeq 1.6487$, and does not change the character of the solutions, as we shall demonstrate in section \ref{num}.

\section{An asymptotic  Green's function solution}\label{sec3}
In this section we derive a global asymptotic Green's function solution to the evolution equation (\ref{gov_pde}). The approach is to write down a single Laplace mode solution which reduces to the solutions of (\ref{gov_pde}) in each of the Newtonian ($r \gg r_I$), near-ISCO ($r \rightarrow r_I$) and global WKB limits.

We begin by defining the dimensionless variables 
\beq
x \equiv {2r \over r_I}, 
\eeq
and 
\beq
h \equiv {J \over \sqrt{GMr_I/2}} = x^{1/2} \exp\left({1\over x}\right),
\eeq
meaning  that our stress parameterisation has the form
\beq
\W = w x^{\mu - 1/2} h . 
\eeq
Writing the governing equation in terms of $h$ and $x$ leaves 
\beq\label{gov}
 \epsilon^2 {{\rm d}^2 y \over {\rm d}h^2}  = - {x^{3/2 - \mu} \over h^2} {1 \over 1 - {2 / x}} y ,
\eeq 
where 
\beq
\epsilon^2 \equiv {w \over r_I s  \sqrt{GMr_I/2}} .
\eeq
This equation remains too complicated to be analytically solved exactly. However, three closed form solutions can be found in physically meaningful limits, as we now demonstrate.  

\subsection{The asymptotic WKB  solution}
A differential equation of the form 
\beq\label{wkb_def}
\epsilon^2 {{\rm d}^2 y \over {\rm d}x^2} = -Q(x) y, \quad Q(x) > 0, 
\eeq
has formal WKB solution, valid in the $\epsilon \rightarrow 0$ limit (Bender \& Orszag 1978), of  
\beq\label{wkb_exp}
y \sim  Q^{-1/4} \cos\left({1 \over \epsilon} \int Q^{1/2} {\rm d}x + \phi\right) ,
\eeq
where $\phi$ is a constant of integration. {It is important to note that the $\epsilon \rightarrow 0$ limit is physically rather natural. As $w$ is defined as the  scale of $\W$, it is of order $w \sim r_I \alpha c_s^2$, where $\alpha \ll 1$ is the usual Shakura-Sunyaev (1973) alpha parameter, and $c_s$ is the disc's speed of sound. Therefore $\epsilon$ is of order
\beq
\epsilon^2 \sim \alpha \left({c_s \over v_K}\right) \left({c_s \over r_I s}\right) ,
\eeq
where we have defined the Keplerian velocity of the ISCO $v_K = \sqrt{GM/r_I} = c \sqrt{r_g/r_I} \gg c_s$. The combination $\alpha (c_s/v_K) \ll 1$, and so $\epsilon \ll 1$, provided that $s$ is not extremely small. Small $s$ modes dominate at very large times (eq. \ref{laplacedef}), and  the suitability of this approximation must be tested numerically. In section \ref{num} we demonstrate that even at large times this approximation holds well. }

The governing equation (\ref{gov}) therefore has the formal WKB solution:
\begin{multline}
y_{\rm WKB} =  x^{1/4} \exp\left({1\over 2x}\right) \left(1-{2\over x}\right)^{1/4} \\ x^{-\alpha/2} \cos\left({1\over \epsilon} f_\alpha (x) + \phi \right),  
\end{multline}
where 
\beq
f_\alpha(x) \equiv \int {[x(h)]^{\alpha} \over h} {1 \over \sqrt{1 - {2 / x(h)}}} \,{\rm d}h , 
\eeq
and 
\beq
\alpha \equiv {3 - 2\mu \over 4}. 
\eeq
After substituting for $h = x^{1/2} e^{1/x}$, this integral becomes 
\beq
f_\alpha(x) = {1\over 2} \int_2^x {x'^{\alpha - 1}  \sqrt{1 - {2 \over  x'}}} \, {\rm d}x' . 
\eeq
By making the substitution $x = 2 \cosh^2(\psi)$, and by repeated integration by parts, this integral can be expressed as a hypergeometric series of $2/x$. The full solution is presented in Appendix  \ref{app}, but the key asymptotic properties of $f_\alpha$ are the following:
\beq
f_\alpha(x \gg 2) \sim {x^\alpha \over 2 \alpha} \left({1 - {2\over x}}\right)^{1/2},
\eeq
and 
\beq
f_\alpha(x \rightarrow 2) \sim  \left(1 - {2\over x}\right)^{3/2}. 
\eeq
It is important to note here that our WKB expression ceases to be a valid asymptotic expansion in the limit $x \rightarrow 2$. This is because, in the notation of equation (\ref{wkb_def}), the function $Q$ is poorly behaved in this region $Q(x\rightarrow 2) \rightarrow \infty$.   Our WKB solution is  not valid near to $x = 2$ and we expect, therefore, that the global solution that we derive should have, in the WKB $\epsilon \rightarrow 0$ limit, leading order corrections  of order $(2/x)$ (i.e., corrections which vanish for large $x \gg 2$, but which are non-negligible near $x=2$).

\subsection{The asymptotic Newtonian  solution}
The Newtonian limit is formally the $x \gg 2$ limit. In this limit we recover the Newtonian angular momentum relationship ($h = \sqrt{x}$), and our governing equation is 
\beq\label{newt}
 \epsilon^2 {{\rm d}^2 y_N \over {\rm d}h^2}  = - {h^{1 - 2\mu}} \, y_N .
\eeq 
This equation is of a standard form 
with known Bessel function solutions (Gradshteyn and Ryzhik {\it et al}. 2007).
Explicitly, the solutions of the Newtonian limit of the disc equation is  
\beq
y_N = x^{1/4} J_{1 \over 4 \alpha} \left({1 \over 2 \alpha \epsilon} x^\alpha\right) .
\eeq
As the Newtonian solution is only valid in the $x \gg 2$ limit, we expect the leading order corrections to our global solution, in the Newtonian limit, to also be of order $(2/x)$. 

\subsection{The asymptotic near-ISCO  solution}
The near-ISCO limit is formally the limit of $x \rightarrow 2$. In this limit the leading order behaviour of the governing disc equation is 
\beq\label{nic}
-s y_I = A {{\rm d} \over {\rm d}x } \left[{1 \over x - 2} {{\rm d}y_I \over {\rm d} x} \right] ,
\eeq
where $A$ is a weak function of $x-2$, and can therefore be treated as a constant. {The solution of this equation is in general a linear superposition of the first derivatives of the Airy and Bairy functions of the negative argument (Abramowitz \& Stegun 1965).     
However, for the particular boundary condition $y_I(r_I) = 0$, it is simpler to instead find the series solution of this equation, as it immediately highlights the asymptotic $x \rightarrow 2$ behaviour of the near-ISCO modal solutions}
\beq
y_I = (x-2)^p \sum_{n = 0}^\infty a_n (x-2)^n . 
\eeq
Direct substitution of the above definition into equation \ref{nic} leads to the constraints 
\beq
- {s \over A} \sum_{n=0}^\infty a_n (x-2)^n = \sum_{n=0}^\infty a_n(n+p)(n+p-2) (x-2)^{n-3} . 
\eeq
By comparing term-by-term the solution of these constraints may be found. The required solution, which vanishes at the location of the ISCO ($x = 2$), has $p =2$, $a_0 \neq 0$, $a_1 = a_2 = 0$, and 
\beq
a_{n+3} = -{s \over A} {1\over (n+3)(n+5)} a_{n}, 
\eeq
and is therefore given by
\beq
y_I(x) = (x-2)^2 \sum_{n=0}^\infty a_{3n} (x-2)^{3n} .
\eeq
The asymptotic near-ISCO behaviour of this solution is 
\beq
y_I \sim \left(1-{2 \over x}\right)^2. 
\eeq
The leading order corrections to the global solution in the near-ISCO limit are  expected to be of order $(1-2/x)^{5}$, which is the next to leading order term in the near-ISCO solution. 

\subsection{The global solution}
We require a global solution $y(x, \epsilon)$ with the following asymptotic properties 
\beq
y(x, \epsilon\rightarrow 0) \sim y_{\rm WKB} + {\cal O}\left[ \left({2 \over x}\right) \right] ,
\eeq
which we shall refer to as the `WKB limit',  
\beq
y(x \rightarrow 2, \epsilon) \sim y_{I} + {\cal O}\left[ \left(1 - {2 \over x}\right)^{5} \right] ,
\eeq
which we shall refer to as the `near-ISCO limit', and
\beq
y(x \gg 2, \epsilon) \sim y_{N}   + {\cal O}\left[ \left({2 \over x}\right) \right] ,
\eeq
which we shall refer to as the `Newtonian limit'. In the above expressions we explicitly display the expected behaviour of the (small) correction terms in each asymptotic limit. 

The solution 
\begin{multline}
y(x, s) = \sqrt{{\pi \over 2 \epsilon} x^{-\alpha} f_\alpha(x)} \exp\left({1\over 2x}\right) \left(1 - {2\over x}\right)^{{5 \over 4} - {3 \over 8\alpha}} \\
x^{1/4} J_{1\over4\alpha}\left({1\over \epsilon }f_\alpha(x)\right),
\end{multline}
satisfies each of these asymptotic behaviours, as we now demonstrate. We begin with the WKB $\epsilon \rightarrow 0$ limit.  The large argument expansion of a Bessel function is the following (Gradshteyn and Ryzhik {\it et al}. 2007):
\beq
\lim_{z \rightarrow \infty} J_\nu(z) = \sqrt{2 \over \pi z} \cos\left(z + \phi_\nu\right) + {\cal O}\left[z^{-3/2}\right] ,
\eeq
where $\phi_\nu = -\pi/4 - \pi\nu/2$ is a constant. 
We therefore have that 
\begin{align}
\lim_{\epsilon\rightarrow 0} y(x, s) &=  x^{1/4} \exp\left({1\over 2x}\right) \left(1-{2\over x}\right)^{1/4} \nonumber \\ &\times x^{-\alpha/2} \cos\left({1\over \epsilon} f_\alpha (x) + \phi \right) \times \left(1 - {2 \over x} \right)^{1 - {3\over8\alpha}} \nonumber \\ &= y_{\rm WKB} + {\cal O}\left[ \left({2 \over x}\right) \right] .
\end{align}
The near-ISCO limit can be treated with the small argument expansion of the Bessel function (Gradshteyn and Ryzhik {\it et al}. 2007):
\beq
\lim_{z \rightarrow 0} J_\nu(z) = z^\nu + {\cal O}\left[z^{\nu + 2}\right],
\eeq
and the $x \rightarrow 2$ limit of $f_\alpha(x\rightarrow 2)\sim  (1-2/x)^{3/2} $: 
\begin{multline}
\lim_{x \rightarrow 2} y(x, s) \sim   \left(1 - {2\over x}\right)^{2 - {3 \over 8\alpha}} \lim_{x \rightarrow 2} \left[ J_{1\over4\alpha}\left(\left(1 - {2 \over x} \right)^{3/2} \right) \right]\\ 
\sim \left(1 - {2 \over x} \right)^2 + {\cal O}\left[ \left(1 - {2 \over x}\right)^{5} \right] = y_I + {\cal O}\left[ \left(1 - {2 \over x}\right)^{5} \right]. 
\end{multline}
Finally, the Newtonian limit follows from the $x \gg 2$ limit of $f_\alpha(x \gg 2) \sim x^\alpha(1-2/x)^{1/2}/2\alpha $:
\begin{multline}
\lim_{x \gg 2} y(x, s) \sim x^{1/4}  J_{1\over4\alpha}\left({x^\alpha\over 2\epsilon\alpha }\left(1 - {2 \over x} \right)^{1/2} \right) \\ \times \exp\left({1 \over 2 x}\right)  \left(1 - {2\over x}\right)^{{3 \over 2} - {3 \over 8\alpha}}  \\  
= x^{1/4} J_{1\over4\alpha}\left({x^\alpha\over 2\epsilon\alpha }\right) + {\cal O}\left[ \left({2 \over x}\right) \right] = y_{N}   + {\cal O}\left[ \left({2 \over x}\right) \right] . 
\end{multline}
\subsection{The Green's function}
We now construct the (unnormalised) Green's function  corresponding to this global modal solution. The superposition integral
\begin{multline}\label{mode_sup}
\int_0^\infty  J_\nu(2\beta \sqrt{s}) J_\nu(2 \gamma \sqrt{s}) e^{-st} {\rm d}s \\
= {1 \over t} \exp\left({-\beta^2 - \gamma^2 \over  t} \right) I_\nu \left({2 \beta \gamma \over  t}\right),
\end{multline}
is of use here  (Gradshteyn and Ryzhik {\it et al}. 2007), as it  approaches a delta function $\delta(\beta - \gamma)$ in the  $t\rightarrow 0$ limit
\beq
\lim_{t\rightarrow 0} {1 \over t} \exp\left({-\beta^2 - \gamma^2 \over  t} \right) I_\nu \left({2 \beta \gamma \over  t}\right) \sim {1\over t^{1/2}} \exp\left(-{(\beta - \gamma)^2 \over t}\right) .
\eeq
Our (unnormalised) asymptotic Green's function is therefore  
\begin{multline}\label{green_y}
G(x, x_0, \tau) = \sqrt{x^{-\alpha} f_\alpha(x) \exp\left({1 \over x} \right) \left(1 - {2\over x}\right)^{5/2 - 3/4\alpha} }\\ 
 {x^{1/4} \over \tau} \exp\left({-f_\alpha(x)^2 - f_\alpha(x_0)^2 \over  4\tau} \right) I_{1\over 4\alpha} \left({ f_\alpha(x) f_\alpha(x_0) \over  2\tau}\right),
\end{multline}
where $x_0$ is the radial location of the initial density spike, and $\tau$ is defined via 
\beq
\tau \equiv \sqrt{2 \over GMr_I} {w t \over r_I} \left(1 - {r_I \over r_0}\right),
\eeq 
where $t$ is measured in physical units. It is important to note that $\tau$ as defined here is {\it not} equal to the time in units of the viscous timescale at the initial radius $\tau \neq t/t_{\rm visc}(r_0)$. 

The Green's function of equation (\ref{green_y}) describes the evolution of the quantity $y = r\Sigma \W$.  It is often  more natural to instead work with the Green's function of $\Sigma$, the disc's surface density. The normalised Green's function for the disc surface density, denoted by $G_\Sigma$, is given by 
\begin{multline}\label{green_s}
G_\Sigma(x, x_0, \tau) = \\ {M_d \over 2\pi r_I^2 c_0}  \sqrt{x^{-\alpha} f_\alpha(x) \exp\left(-{1 \over x} \right) \left(1 - {2\over x}\right)^{5/2 - 3/4\alpha} }\\ 
 {x^{-3/4 - \mu} \over \tau} \exp\left({-f_\alpha(x)^2 - f_\alpha(x_0)^2 \over  4\tau} \right) I_{1\over 4\alpha} \left({ f_\alpha(x) f_\alpha(x_0) \over  2\tau}\right),
\end{multline}
where 
\beq
c_0 = x_0^{(1 + 14\mu)/8} \left(1 - {2\over x_0}\right)^{3/4 - 3/8\alpha} \exp\left(-{1\over 2x_0}\right) \sqrt{1\over f_\alpha(x_0)} ,
\eeq
and $M_d$ is the total mass of the disc at $t = 0$. This normalisation constant was found by computing the integral 
\beq
\int_{r_I}^\infty 2\pi r G_\Sigma(x, x_0, \tau \rightarrow 0) \, {\rm d} r = M_d ,
\eeq
and by using the following delta function identity 
\begin{align}
\delta\left(f_\alpha(x) - f_\alpha(x_0)\right) &= {1 \over \left| \partial_x f_\alpha(x) \right|_{x_0}} \, \delta(x - x_0) , \nonumber \\ &= {2x_0^{1-\alpha} \over \sqrt{1 - 2/x_0}} \, \delta(x - x_0) .
\end{align}

\section{Numerical verification}\label{num}
The numerical test of this theory is simple.   We compute the numerical Green's function solutions of equation (\ref{27q}), the full general relativistic disc equation,  for a variety of different values of the Kerr spin parameter, and compare them to the the properties of the analytical solutions  (equation \ref{green_y}). For this test we take $\mu = 0$ in the stress parameterisation:  the numerical solutions where found for 
\beq
\W = w = {\rm cst}.
\eeq
The boundary condition imposed was that of a vanishing ISCO stress, by which we mean that $\zeta(r_I)$ was set to zero throughout the simulation. With the numerical solutions of $\zeta(r, t)$ found, the combination $r\Sigma \W$ was extracted, so that it could be compared with the Green's function solution for $y$. 

The analytical solutions are given by equation (\ref{green_y}). For stress parameterisation  $\mu = 0$ we have $\alpha = 3/4$, and the function $f_{\alpha}(x)$ has the following exact form 
\begin{multline}
f_{3\over4}(x) = {2 \over 3}x^{3/4}  \sqrt{1 - {2\over x}}\left[1  + {4 \over x}\, {}_2F_1\left(1, {3\over 4}; {5\over 4}; {2\over x}\right) \right] \\ - {2^{11/4} \over3}\sqrt{\pi} {\Gamma(5/4)  \over  \Gamma({3/4})} ,
\end{multline}
where ${}_2 F_{1}$ is the hypergeometric function and $\Gamma$ is the gamma function (see Appendix \ref{app} for a full discussion and derivation of this solution). The (arbitrary) normalisations of the numerical and analytical solutions where chosen so that they had the same peak amplitude at $t/t_{\rm visc} = 0.1$.

\begin{figure}
\includegraphics[width=\linewidth]{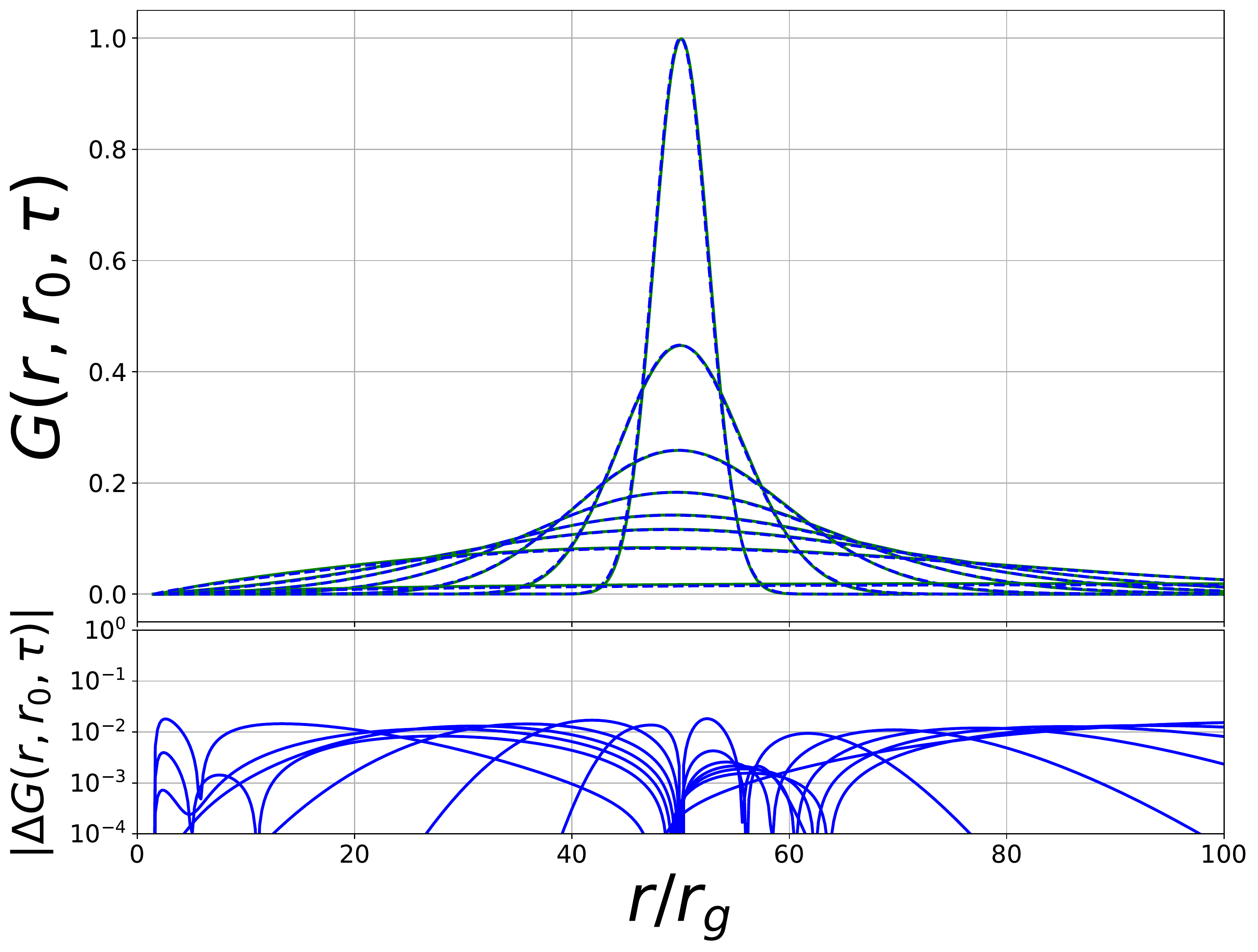}
\caption{Upper: The Green's function solution of the variable $y \equiv r \Sigma \W$, for a Kerr black hole with spin $a = +0.99$. The blue dashed curves are the numerical solutions of the full general relativistic disc equations, while the green solid curves are the analytical solution of equation \ref{green_y}. The initial radius was $r_0 = 50r_g$ and the curves are plotted at dimensionless times $t/t_{\rm visc} = 0.003, 0.015, 0.045, 0.09, 0.15, 0.225, 0.45$ and $4.5$.  Lower: The absolute value of the difference between the numerical and analytical Green's function solutions.  To allow a proper comparison at each time both the numerical and analytical Green's functions are renormalised to have a peak amplitude of $1$ (see equation \ref{delta_def}).      }
\label{max_spin_test}
\end{figure}

\begin{figure}
\includegraphics[width=\linewidth]{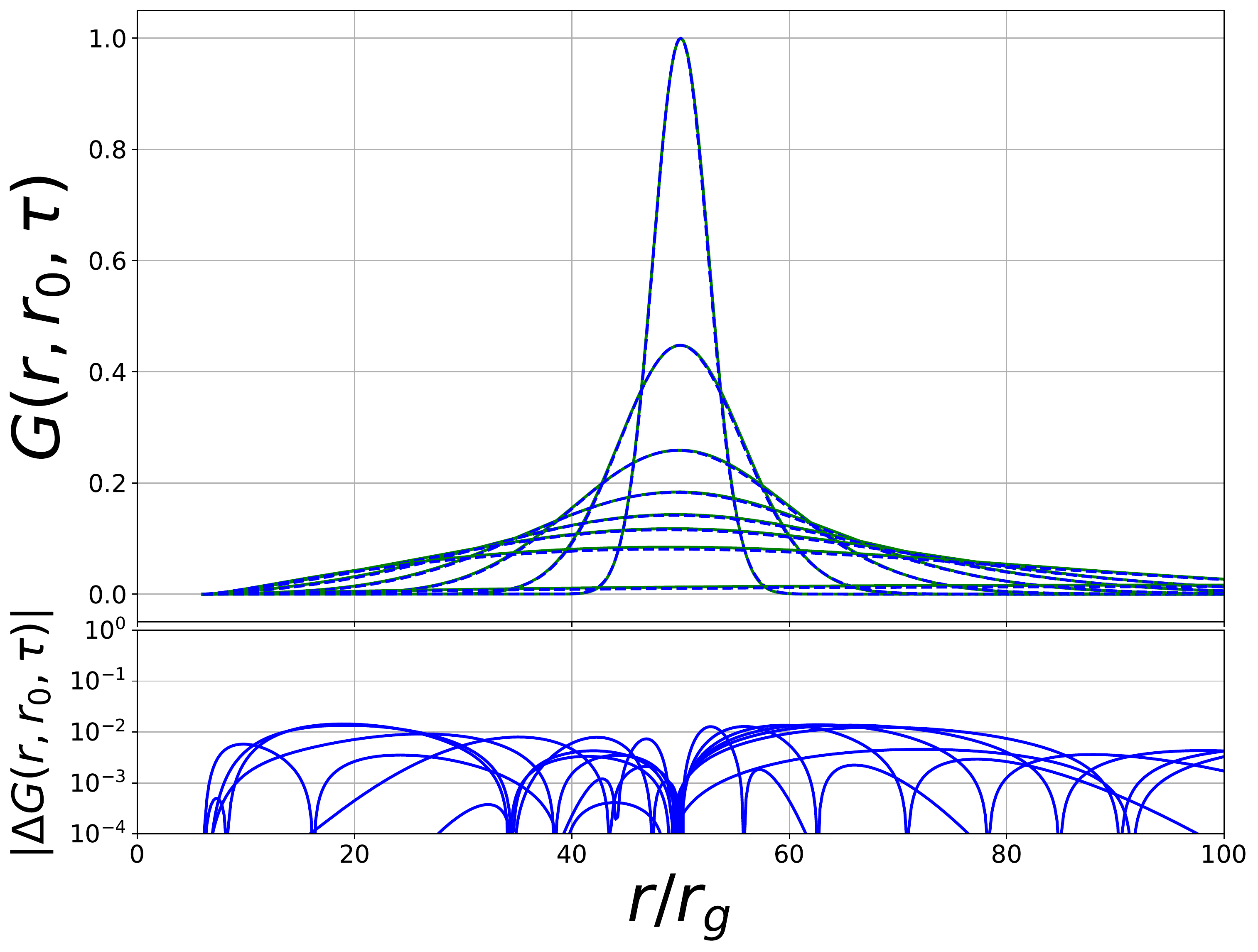}
\caption{Same as Figure \ref{max_spin_test}, but for $a = 0$. }
\label{zero_spin_test}
\end{figure}

\begin{figure}
\includegraphics[width=\linewidth]{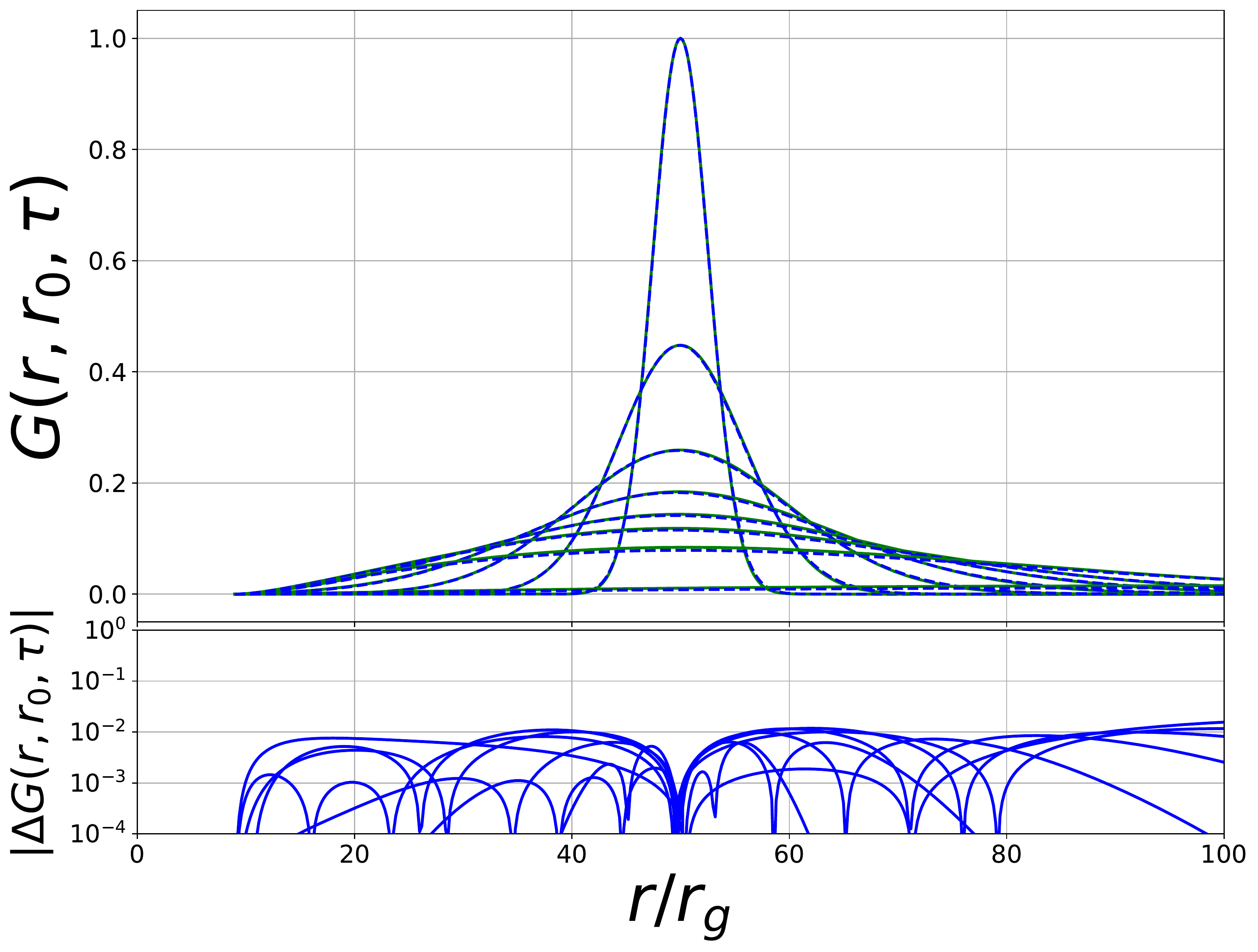}
\caption{Same as Figure \ref{max_spin_test}, but for $a = -0.99$. }
\label{minus_max_spin_test}
\end{figure}

In Figure \ref{max_spin_test} we plot the numerically (blue dashed curve) and analytically (green solid curve) computed $r\Sigma \W$ profiles, assuming an initial radius of $r_0 = 50r_g$ and Kerr angular momentum parameter $a = +0.99$.  The curves are plotted at dimensionless times $t/t_{\rm visc} = 0.003, 0.015, 0.045, 0.09, 0.15, 0.225, 0.45$ and $4.5$, the curves at later times are identifiable through their decreasing peak amplitude.  It is remarkable how accurately the analytical Green's function solution of equation (\ref{green_y}) reproduces the properties of the full numerical solutions. {This is true even at the latest times, when the WKB approach is formally invalid. In appendix \ref{obvious_stuff} we contrast this behaviour with the Newtonian Green's function solutions (Lynden-Bell \& Pringle 1974), which poorly reproduce the properties of the numerical relativistic solutions.  } 
 
 A Kerr spin parameter of $a = +0.99$ is precisely the parameter regime where it might be expected  that the approximations used in constructing this asymptotic solution would be least accurate, as the spin dependent terms in the relativistic angular momentum gradient (eq. \ref{rel_ang_mom}) appear substantial in this limit.  However, at least for the case of a vanishing ISCO stress, this is not the case. The Green's function solution of the preceding section remain extremely accurate for all values of the Kerr black hole spin parameter.  This result is further emphasised in Figures \ref{zero_spin_test} and \ref{minus_max_spin_test}. In Figures \ref{zero_spin_test} and \ref{minus_max_spin_test} we display identical calculations as in Figure \ref{max_spin_test}, but for spin parameters $a = 0$ (Fig. \ref{zero_spin_test}) and $a = -0.99$ (Fig. \ref{minus_max_spin_test}). {It is interesting to note that the asymptotic  solution of the proceeding section is independent of the black hole's spin, once the disc radius is normalised by the ISCO value. This is a property also exhibited by the full numerical solutions of the relativistic disc equations, as verified in Appendix \ref{spin_stuff}.}

The absolute discrepancies between the numerical and analytical solutions, for each of Figs. \ref{max_spin_test},  \ref{zero_spin_test} and \ref{minus_max_spin_test} are all at the sub-percent level.  To verify this, we compute (where $G_{\rm num}$ is the numerical solution, and $G$ the analytical solution)
\beq\label{delta_def}
\left| \Delta G(x, x_0, \tau) \right| \equiv \left| {G_{\rm num}(x, x_0, \tau) \over \max\left[ G_{\rm num}(x, x_0, \tau) \right]}- {G(x, x_0, \tau) \over \max\left[ G(x, x_0, \tau) \right] } \right| ,
\eeq
at each time (lower panels of Figs.  \ref{max_spin_test}, \ref{zero_spin_test}, \ref{minus_max_spin_test}). We confirm that for all radii and times shown in Figs. \ref{max_spin_test} -- \ref{minus_max_spin_test} this quantity satisfied  $\left| \Delta G(x, x_0, \tau) \right| \lesssim 0.01$. This finding was independent of the black hole spin parameter used.

\section{The Green's function for the mass accretion rate}\label{sec5}
For some applications it is more convenient to work with the Green's function of the mass accretion rate, rather than the surface density or variable $y$ (e.g., Mushtukov et al. 2018).   The mass accretion rate within the disc is given by 
\beq
\dot M(r, t) = 2\pi r u^r \Sigma ,
\eeq
where $\Sigma$ is the disc surface density, and $u^r$ the accretion flows radial velocity.  For a Newtonian disc, the radial velocity of the accretion flow is given by
\beq
u^r = - {\W \over J'} {1 \over y} {\partial y \over \partial r},
\eeq
and therefore
\beq
\dot M = - {2\pi  \over J'} {\partial  y \over \partial r} = - 2\pi \sqrt{r \over GM} {\exp\left(-{r_I / 2 r}\right) \over 1 - r_I/r} {\partial y \over \partial r} ,
\eeq
for our pseudo-potential. To compute the mass accretion rate the gradient of our Green's function must be found.   We shall define 
\beq 
G_{\dot M} (x, x_0, \tau) \equiv - {x^{1/2} \exp\left(-1/x\right)  \over 1 - 2/x} {\partial  \over \partial x} G(x, x_0, \tau),
\eeq
which is the (unnormalised) Green's function of the mass accretion rate.   Taking the derivative of the asymptotic Green's function is complicated slightly by the various relativistic correction factors, and a full treatment is presented in Appendix \ref{gradient}. An important identity used in this analysis is the following (Gradshteyn and Ryzhik {\it et al}. 2007)
\beq
{{\rm d} \over {\rm d}z} I_l(z) = I_{l-1}(z) - {l \over z} I_l(z).
\eeq
The (unnormalised) Green's function of the mass accretion rate is the following 
\begin{multline}\label{green_m}
G_{\dot M}(x, x_0, \tau) = - \sqrt{x^{\alpha} f_\alpha(x) \exp\left(-{1 \over x}\right) \left(1 - {2\over x}\right)^{1/2 - 3 / 4\alpha}} \\
{x^{3/ 4}\over \tau} \exp\left({-f_\alpha(x)^2 - f_\alpha(x_0)^2 \over  4\tau} \right) \\ 
\Bigg[ \Bigg(P(x, \alpha) - {f_\alpha(x) \over 4 \tau} x^{\alpha-1} \sqrt{1 - {2\over x}} \Bigg) I_{1\over 4\alpha} \left({ f_\alpha(x) f_\alpha(x_0) \over  2\tau}\right)  \\
+  \left( {f_\alpha(x_0) \over 4 \tau} x^{\alpha-1} \sqrt{1 - {2\over x}} \right)  I_{1 - 4\alpha \over 4\alpha} \left({ f_\alpha(x) f_\alpha(x_0) \over  2\tau}\right) \Bigg] ,
\end{multline}
where $P(x, \alpha)$ contains all the explicit small $x$ relativistic  corrections terms. This function satisfies $P(x\gg 2, \alpha) \rightarrow 0$ and is given explicitly  by 
\begin{multline}
P(x, \alpha) = {1 - 2\alpha \over 4x} - {1 \over 2x^2} + {10\alpha - 3 \over 4\alpha x^2}{1 \over 1 - 2/x}\\ + {1 \over 4}\left(1 - {1 \over \alpha} \right) {x^{\alpha - 1}\sqrt{1-2/x} \over f_\alpha(x)}.
\end{multline}

\begin{figure}
\includegraphics[width=\linewidth]{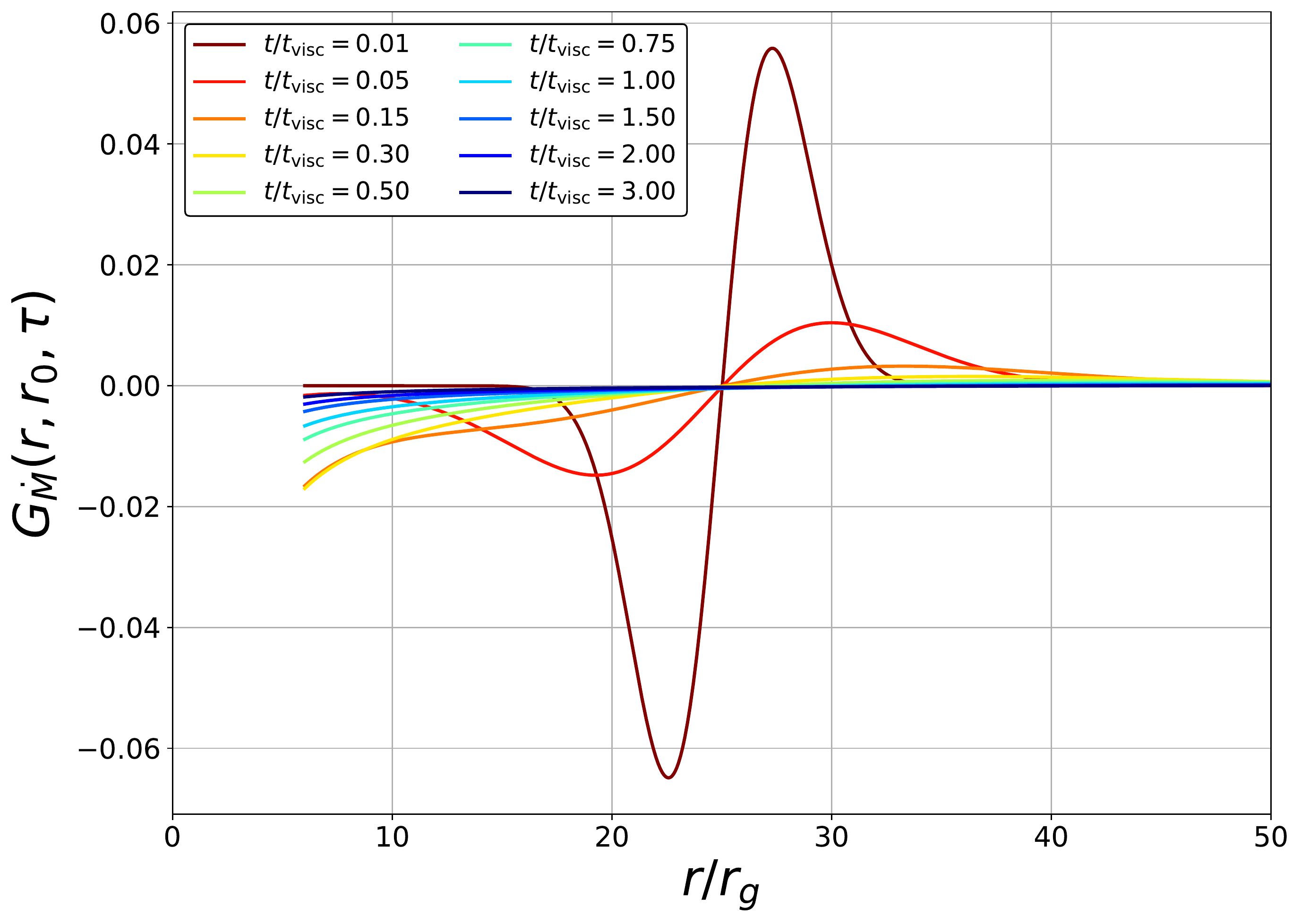}
\caption{The Green's function solution of the mass accretion rate for a Schwarzschild black hole  ($a = 0$). The initial radius was $r_0 = 25r_g$ and the curves are plotted at the dimensionless times denoted in the legend.  The normalisation of the accretion rate was chosen so that the time-integrated ISCO accretion rate was equal to 1.  }
\label{zero_spin_mac}
\end{figure}

\begin{figure}
\includegraphics[width=\linewidth]{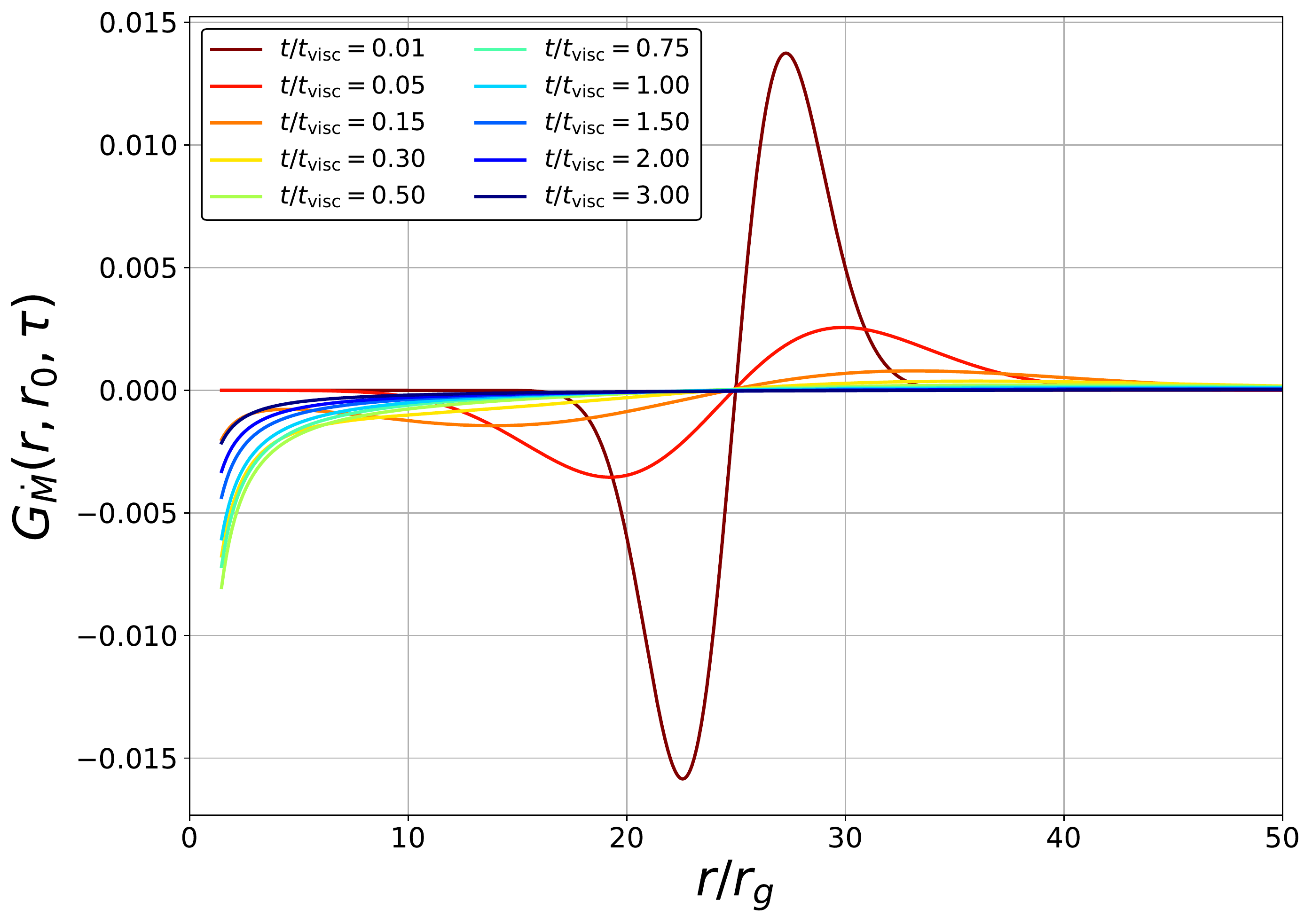}
\caption{Same as Figure \ref{zero_spin_mac}, but for $a = +0.99$. Note that the radial accretion rate is more sharply peaked towards the ISCO for higher black hole spins.  }
\label{max_spin_mac}
\end{figure}

In Figures \ref{zero_spin_mac} and \ref{max_spin_mac} we show the mass accretion rate Green's function, as a function of radius, for the spin parameters $a = 0$ (Fig. \ref{zero_spin_mac}) and $a = +0.99$ (Fig. \ref{max_spin_mac}), at a number of different dimensionless times denoted on each plot. The initial radius in both cases was taken to be $r_0 = 25r_g$.  A value of $\dot M(r, t) < 0$ denotes mass inflow (towards the ISCO), while $\dot M(r, t) > 0$ denotes mass outflow. Some mass must move outwards within the disc so as to conserve the total angular momentum of the flow.  The normalisation of the accretion rate in both plots was chosen so that the time-integrated ISCO accretion rate was equal to 1. A result of some interest is that the radial mass accretion rate is more sharply peaked towards the ISCO for larger black hole spins.

The mass accretion rate across the ISCO is a quantity  which can be described rather simply analytically. Computing the ISCO accretion rate involves taking the $x \rightarrow 2$ limit of $G_{\dot M}(x, x_0, \tau)$.   Taking this limit we find that the various factors of $(1-2/x)$ all cancel, and the ISCO accretion rate takes the following  form 
\beq
G_{\dot M}(x_I, x_0, \tau) = {C_\alpha \over \tau^{1 + 1/4\alpha}} \exp\left(-{f_\alpha(x_0)^2 \over 4 \tau}\right) ,
\eeq
where $C_\alpha<0$ is a constant. It is of interest to normalise the ISCO accretion rate by its maximum value, which occurs at a dimensionless  time 
\beq
\tau_{\rm peak} = {f_\alpha(x_0)^2 \over 4 + 1/\alpha} .
\eeq
In doing so  the ISCO accretion rate is given by 
\beq\label{macico}
G_{\dot M}(x_I, x_0, \tau) = {{\dot M}_{\rm peak} \over \tau^n} \left({e f_a(x_0)^2 \over 4n}\right)^n  \exp\left(-{f_\alpha(x_0)^2 \over 4 \tau}\right) ,
\eeq
where $n \equiv 1 + 1/4\alpha$. The late-time behaviour of the disc luminosity is well approximated by the behaviour of the ISCO accretion rate (i.e.,  proportional to $t^{-n}$), a result which is in keeping with the numerical analysis of Balbus \& Mummery (2018). 

The normalisation of the ISCO accretion rate can be related to the normalisation of the Green's function for the surface density by noting that the entirety of the disc material is eventually accreted. Setting 
\beq
\int_0^\infty G_{\dot M}(x_I, x_0, \tau) \, {\rm d}t = - M_d, 
\eeq
(where $t$ is the  time in physical units) we find that 
\beq
\dot M_{\rm peak} = - \sqrt{2 \over GMr_I} {M_d w (1-r_I/r_0) \over r_I f_\alpha(x_0)^2} \left({4n \over e}\right)^n {2^{-1/2\alpha} \over \Gamma(1/4\alpha)} .
\eeq

\begin{figure}
\includegraphics[width=\linewidth]{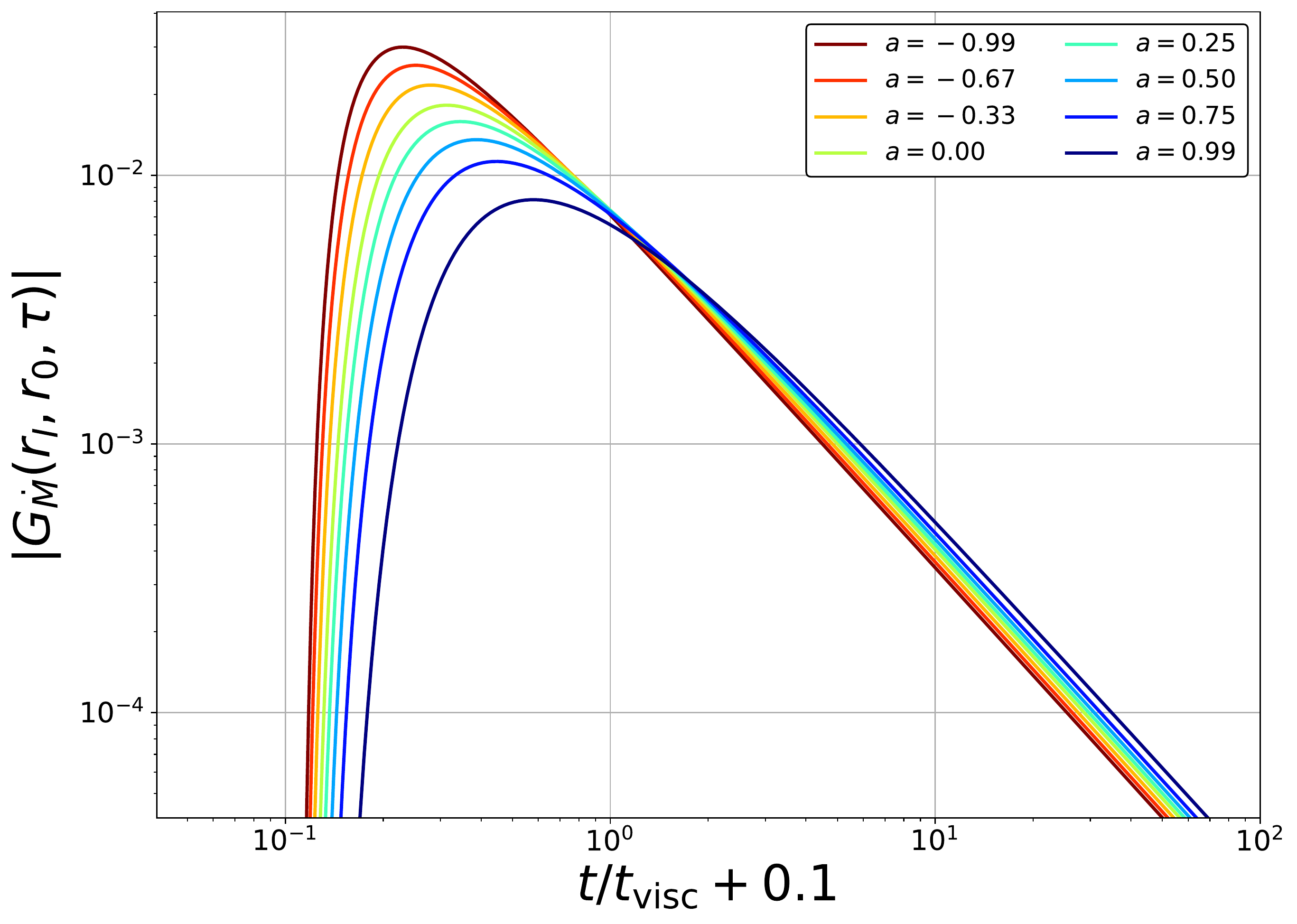}
\caption{The ISCO accretion rate as a function of time, for Kerr black holes of different spins (displayed in legend). The time integrated ISCO accretion rate of each curve is equal to 1.    }
\label{isco_acc_spin}
\end{figure}

The absolute value of the ISCO accretion rate, plotted as a function of time, is shown in Figure \ref{isco_acc_spin} for a number of different values of the Kerr spin parameter $a$. Each curve is normalised so that the  time integrated ISCO accretion rate is equal to 1.  The initial radius of each curve was $r_0 = 25r_g$.  It is interesting to note that the absolute value of the peak ISCO accretion rate, for constant disc mass, $w$,  and $r_0$, {\it decreases} as a function of increasing spin.  Conversely, relative to their peak value, the ISCO accretion rates of more rapidly rotating black holes are larger for considerably {\it longer} than their more slowly spinning counterparts. 

While the peak values of $\dot M(r_I)$ are  lower for higher black hole spins, it is important to remember that  discs around more rapidly spinning Kerr black holes have higher mass-to-light efficiencies. Therefore, for a given  ISCO accretion rate, the rest-frame luminosity of a more rapidly rotating Kerr black hole will be higher, as will their total integrated energies.

\begin{figure}
\includegraphics[width=\linewidth]{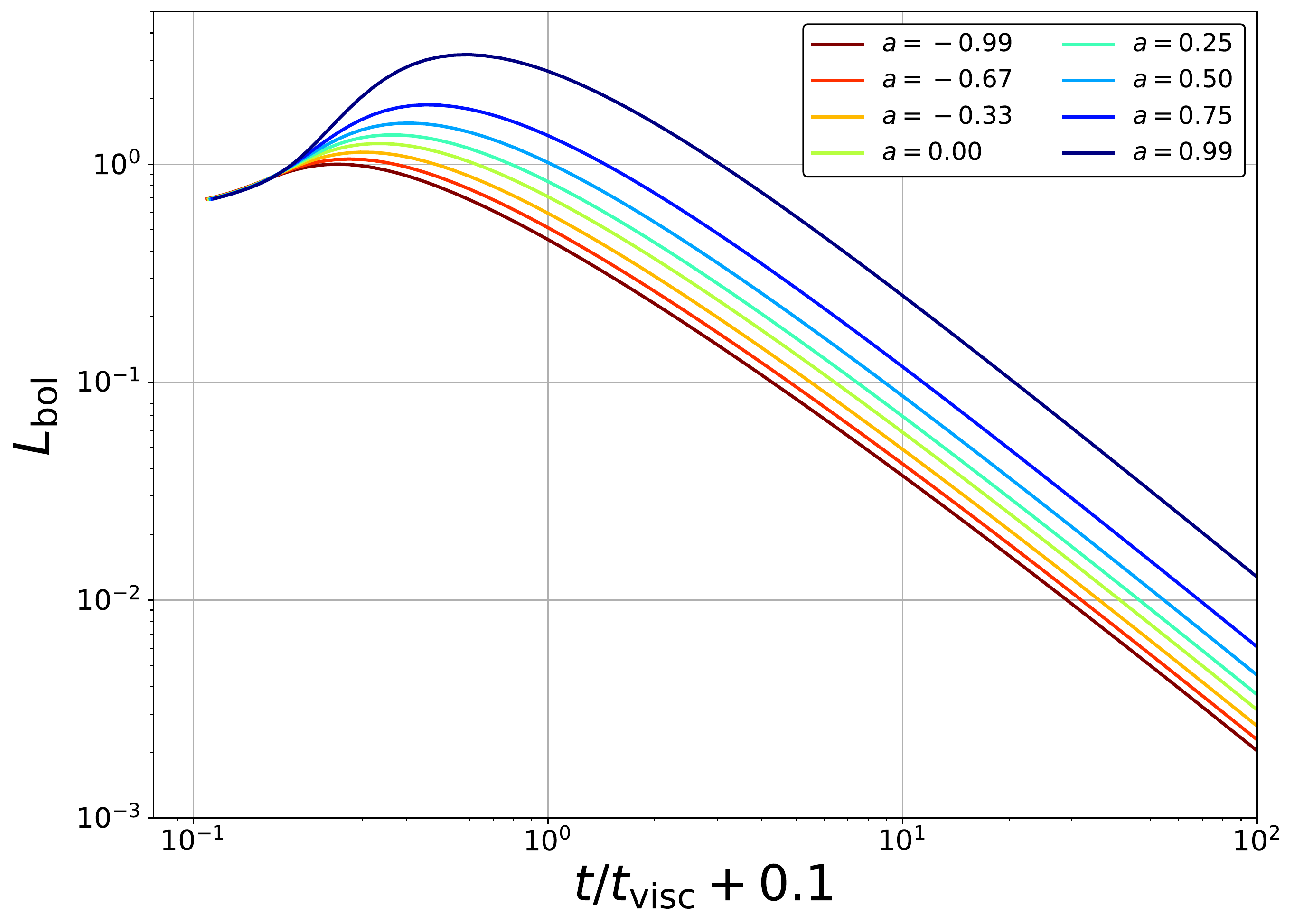}
\caption{The disc bolometric luminosity (defined in text), plotted as a function of time for Kerr black holes of different spins (displayed in legend). The initial disc mass of each curve is set equal to 1.    }
\label{bol_spin}
\end{figure}

This can be seen more clearly in Figure \ref{bol_spin}, where we plot the bolometric luminosities of the discs evolving in Fig. \ref{isco_acc_spin}, as a function of time. The bolometric luminosity is found by integrating the local quantity  $\sigma T^4$ over the entire area of the disc.  A factor of $(1/U^0)^2$ is included to incorporate the effects of the gravitational redshifting of the emitted disc photons. The light curves are then normalised by the peak value of the $a = -0.99$ luminosity.  Each disc had the same initial disc mass.  It is clear to see that the bolometric luminosity of the discs around Kerr black holes increases as a function of the Kerr spin parameter $a$. 

\section{Discussion}\label{sec6}
In this paper we have derived and analysed the leading order Green's function solutions of the  general relativistic thin disc equations, assuming a boundary condition of a vanishing ISCO stress. These solutions promise to be of great practical utility to the astronomical community. 

{It is important to remember, however, that the solutions derived in this paper assume that the turbulent stress in the disc is described by a power-law with radius, and that the disc equations are therefore linear. Many models of the turbulent stress in accretion discs (e.g., Shakura \& Sunyaev 1973) lead to non-linear evolutionary equations, which may well be more physically motivated. This simplification should be kept in mind when modelling astronomical sources with the solutions derived here. } In addition, a natural extension of this model  would be to include non-zero stresses at the discs inner edge. This transpires to be a non-trivial extension, as we discuss below.  

\subsection{The case of a non-zero ISCO stress}
The solutions of the relativistic disc equations, for discs with a non-zero ISCO stress, exist on a continuum of solutions with properties which depend on the magnitude of the stress at the inner edge (Mummery \& Balbus 2019). In the outer Newtonian regions, these solutions are given by a superposition of $J_{+1/4\alpha}$ and $J_{-1/4\alpha}$ modes, with relative amplitudes determined by the inner-disc properties.  At large times the solutions for any finite ISCO stress are eventually dominated by the positive index vanishing ISCO stress modes discussed in this paper, after a transient phase where the negative index modes dominate. The length of this transitional phase is controlled by the magnitude of the inner disc stress, ranging from zero time (a completely vanishing ISCO stress) to a formally infinite length of time (a heavily stressed solution; see Mummery \& Balbus 2019 for more details). 

The modal techniques used in this paper could in principal be extended to incorporate a modal admixture in the Newtonian regions, and a modified inner boundary condition. However, no simple analytical solution of the modal superposition integral (eq. \ref{mode_sup}) exists for the case of a mixture of Bessel functions, and  these solutions therefore lose their simple closed form properties. In addition, discs with finite ISCO stresses have significant densities in the $x \rightarrow 2$ region, where our WKB modal solutions become invalid.  For these reasons it seems likely that numerical techniques will remain the best avenue for studying relativistic discs with finite ISCO stresses.   

\subsection{Implications and applications}
In addition to deriving the leading order Green's function solutions of the  general relativistic thin disc equations, in this paper we have analysed a number of their key properties.  Of particular interest is a simple expression for the time-dependent accretion rate across the ISCO (eq. \ref{macico}).  One implication of this result is that the late-time behaviour of the ISCO mass accretion rate is, for any initial condition,  equal to that of the $r=0$  accretion rate in a Newtonian disc.  This can be seen by noting that for general initial condition $\Sigma(r_0)$, the ISCO accretion rate is 
\begin{align}
\dot M(r_I, t) &= \int_{r_I}^\infty G_{\dot M}(x_I, x_0, \tau) \Sigma(r_0) \, {\rm d}r_0, \nonumber \\ 
&\propto t^{-n} \int_{r_I}^\infty [f_\alpha(x_0)]^{2n} \Sigma(r_0) \exp\left(-{f_\alpha(x_0)^2 \over 4 \tau} \right) \, {\rm d}r_0, \nonumber \\ 
&\sim t^{-n}, \quad t \rightarrow \infty, \quad n \equiv 1+ 1/4\alpha = {4-2\mu \over 3-2\mu}.
\end{align}

We have demonstrated that the Kerr black hole spin parameter $a$ influences  the inner-disc accretion rate in three key ways.  The accretion rate for a rapidly rotating Kerr black hole is more sharply peaked in a narrow radial region around the ISCO.    The peak mass accretion rate across the ISCO  is, for fixed disc mass, $w$ and $r_0$, {\it lower} for more rapidly rotating Kerr black holes. However, the ISCO accretion rate remains higher as a fraction of its peak value for much {\it longer} for these rotating black holes.  When the differing mass-to-light efficiencies of discs in different Kerr spacetimes are taken into account, the rest-frame luminosities of the more rapidly spinning black holes are largest.

The most interesting astrophysical application of this work is likely to be  the study of time-dependent emission from black hole discs. Systems of potential interest include the dramatic state changes observed in many black hole X-ray binary systems. It is thought that the inner regions of these discs  disappear and then reform during these events, and these solutions may aid in the study of this reformation process.   In addition, the numerical solutions of the relativistic disc equations have been of practical utility in studying the evolution of tidal disruption event discs,  which form from the debris of a mangled (formerly normal) star. The development of accurate analytical solutions of these equations will aid in the future data analysis of these remarkable sources.   

Finally, these solutions will be of  utility for the study of aperiodic variability in black hole sources.   Many properties of the observed variability of black hole accretion disc sources are rather naturally explained by the so-called `theory of propagating  fluctuations'  (Lyubarskii 1997), in which fluctuations arise at different radial coordinates of the accretion disc and then propagate towards the central object. Each perturbation of the discs accretion rate can be described by its own Green's function solution.   Heretofore the analysis of this theory has utilised the Newtonian Green's functions (see e.g., Mushtukov {\it et al}. 2018), this theory  can now be re-analysed with a `relativistic upgrade'.   

\section*{Acknowledgments} 
I would like to thank  A. Mushtukov for interesting discussions which initiated this work, and S. Balbus for his encouragement.  

\section*{Data accessibility statement}
No   data was used in producing this manuscript.

\appendix{}
\section{The solution of the $f_\alpha(x)$ integral} \label{app}
\subsection{General solution}
We wish to solve the integral 
\beq
f_\alpha(x) = {1\over 2} \int_2^x x'^{\alpha - 1} \sqrt{1 - {2 \over x'}} \, {\rm d} x'. 
\eeq
Begin by making the substitution 
\beq
x' = 2 \cosh^2(\psi) ,
\eeq
which transforms the integral into the form 
\beq
f_\alpha = 2^\alpha \int_0^{\psi^*(x)} \sinh^2 \psi \cosh^{2\alpha-2} \psi \, {\rm d}\psi ,
\eeq
where 
\beq
\psi^*(x) = \arccosh\sqrt{x/2} .
\eeq
A useful result is the following 
\beq
\sinh \left(\arccosh \sqrt{x/2} \right)= \sqrt{{x\over2} - 1} .
\eeq
Assuming that $\alpha \neq 0$ (special cases will be considered later), we can use the reduction formula (Gradshteyn and Ryzhik {\it et al}. 2007)
\begin{multline}
\int \sinh^p \psi \cosh^q \psi \, {\rm d}\psi = {\sinh^{p-1} \psi \cosh^{q+1}\psi \over p + q} \\ - {p-1 \over p + q} \int \sinh^{p-2} \psi \cosh^{q} {\rm d}\psi
\end{multline}
to write this as
\beq
f_\alpha = {x^\alpha \over 2 \alpha} \sqrt{1 - {2\over x}} - {2^{\alpha} \over 2 \alpha}\int_0^{\psi^*}  \cosh^{2\alpha-2} \psi \, {\rm d}\psi ,
\eeq
where we note that the second integral is of order $(2/x)$ smaller than the previous integral.   The second integral can now be reduced using (Gradshteyn and Ryzhik {\it et al}. 2007)
\begin{multline}
\int \sinh^p \psi \cosh^q \psi \, {\rm d}\psi = {\sinh^{p+1} \psi \cosh^{q-1}\psi \over p + q} \\ + {q - 1 \over p + q} \int \sinh^{p} \psi \cosh^{q-2} {\rm d}\psi
\end{multline}
with $p=0$ to leave 
\begin{multline}
f_\alpha = {x^\alpha \over 2 \alpha} \sqrt{1 - {2\over x}}\left[1 - {1\over 2\alpha - 2} \left({2 \over x}\right)\right] \\ - {2^{\alpha} \over 2 \alpha}{2\alpha - 3 \over 2 \alpha - 2} \int_0^{\psi^*}  \cosh^{2\alpha-4} \psi \, {\rm d}\psi .
\end{multline}
Which is again of the same form as the previous equation, with the remaining integral another order of $(2/x)$ smaller. In the general case where $\alpha \not\in \mathbb{Z}^+$, this procedure can be repeated an infinite number of times to leave 
\beq
f_\alpha = {x^\alpha \over 2 \alpha} \sqrt{1 - {2\over x}} \left[ 1 - {1\over 2\alpha - 1} \sum_{k = 0}^\infty \beta_k \left({2\over x}\right)^{k+1}\right] + {\cal R}_\alpha,
\eeq
where 
\beq
\beta_k = {(2\alpha - 1)(2\alpha - 3)\dots (2\alpha - (2k+1)) \over (2\alpha - 2)(2\alpha - 4)\dots (2\alpha - (2k + 2))} ,
\eeq
and ${\cal R}_\alpha$ is a constant of integration defined so that $f_\alpha(x = 2) = 0$. The coefficients $\beta_k$ are a rational function of $k$, meaning that this result can be expressed as a hypergeometric function. This is of practical interest as the hypergeometric functions are cataloged functions with known identities which will assist in computing ${\cal R}_\alpha$. In addition, the hypergeometric functions are implemented in common programming languages, simplifying the numerical implementation of $f_\alpha(x)$ and of the Green's function solutions derived in this paper.

We write the summation out as 
\begin{multline}
f_\alpha = {x^\alpha \over 2 \alpha} \sqrt{1 - {2\over x}} + {\cal R}_\alpha - {x^{\alpha - 1} \over {2\alpha (\alpha - 1)}} \sqrt{1 - {2\over x}} \\ \times \Big[ 1 + {(3/2 - \alpha) (1) \over (2 - \alpha)} {1 \over 1 !} \left({2\over x}\right) \\ + {(3/2 - \alpha)(3/2 - \alpha + 1) (1)(1+1) \over (2 - \alpha) (2-\alpha + 1)} {1 \over 2 !} \left({2\over x}\right)^2 + \dots \Big]
\end{multline}
where the terms in the final bracket are written in such a way so as to reproduce the definition of the hypergeometric function (Gradshteyn and Ryzhik {\it et al}. 2007)
\beq
{}_2F_1(a,b ;c ;z) \equiv 1 + {a b \over c}{z\over 1!} + {a(a+1)b(b+1)\over c(c+1)} {z^2 \over 2!} +\dots
\eeq
thus our solution is 
\beq
f_\alpha = {x^\alpha \over 2 \alpha} \sqrt{1 - {2\over x}}\left[1  - {x^{ - 1} \over { (\alpha - 1)}} {}_2F_1\left(1, {3\over 2}-\alpha; 2-\alpha; {2\over x}\right) \right] + {\cal R}_\alpha.
\eeq
With the solution written in terms of the hypergeometric function, we can compute the constant of integration ${\cal R}_\alpha$. Using (Gradshteyn and Ryzhik {\it et al}. 2007)
\begin{multline}
{}_2F_1(a, b ;c ; z) = {\Gamma(c)\Gamma(c-a-b) \over \Gamma(c-a)\Gamma(c-b) } {}_2F_1(a, b ; a+b + 1 -c; 1-z) \\
+ {\Gamma(c)\Gamma(a + b - c) \over \Gamma(a)\Gamma(b) } (1-z)^{c-a-b}  {}_2F_1(c-a, c-b ; c + 1 - a - b; 1-z) ,
\end{multline}
we can take the $x \rightarrow 2$ limit of our expression to recover ${\cal R}_\alpha$. Explicitly, using ${}_2F_1(a, b; c; 0) = 1$, we have 
\begin{multline}\label{limit}
\lim_{x\rightarrow 2} \left( \sqrt{1-{2\over x}}  {}_2F_1\left(1, {3\over 2}-\alpha; 2-\alpha; {2\over x}\right)\right) = \\ \lim_{x\rightarrow 2} \left(\sqrt{1- {2\over x}} {\Gamma(2-\alpha) \Gamma(-1/2) \over \Gamma(1-\alpha)\Gamma(1/2) }  
+ {\Gamma(2-\alpha) \Gamma(1/2) \over \Gamma(1) \Gamma(3/2 - \alpha)} \right) \\ = \sqrt{\pi} {\Gamma(2-\alpha)  \over  \Gamma({3/ 2} - \alpha)} ,
\end{multline}
and therefore 
\beq
{\cal R}_\alpha = {2^{\alpha - 2} \over \alpha (\alpha - 1)}\sqrt{\pi} {\Gamma(2-\alpha)  \over  \Gamma({3/ 2} - \alpha)}. 
\eeq
This solution of our integral is therefore 
\begin{multline}
f_\alpha(x) = {x^\alpha \over 2 \alpha} \sqrt{1 - {2\over x}}\left[1  - {x^{ - 1} \over { (\alpha - 1)}} {}_2F_1\left(1, {3\over 2}-\alpha; 2-\alpha; {2\over x}\right) \right] \\ + {2^{\alpha - 2} \over \alpha (\alpha - 1)}\sqrt{\pi} {\Gamma(2-\alpha)  \over  \Gamma({3/ 2} - \alpha)}. 
\end{multline}
\subsection{Special cases}
It is easiest to analyse the special cases of the integral $f_\alpha$ by writing the solution in its explicit summation form 
\beq
f_\alpha = {x^\alpha \over 2 \alpha} \sqrt{1 - {2\over x}} \left[ 1 - {1\over 2\alpha - 1} \sum_{k = 0}^\infty \beta_k \left({2\over x}\right)^{k+1}\right] + {\cal R}_\alpha,
\eeq
where 
\beq
\beta_k = {(2\alpha - 1)(2\alpha - 3)\dots (2\alpha - (2k+1)) \over (2\alpha - 2)(2\alpha - 4)\dots (2\alpha - (2k + 2))} .
\eeq

\subsubsection{$\alpha = 0$}
When $\alpha = 0$ we must return to the very first hyperbolic integral 
\beq
f_\alpha = 2^\alpha \int_0^{\psi^*} \sinh^2 \psi \cosh^{2\alpha-2} \psi \, {\rm d}\psi .
\eeq
which becomes 
\beq
f_0 = \int_0^{\psi^*} \tanh^2 \psi  \, {\rm d}\psi  = \psi^* - \tanh\psi^*
\eeq
and therefore
\beq
f_0(x) = - \sqrt{1-{2\over x}} + \arccosh\sqrt{x\over 2} 
\eeq

\subsubsection{$\alpha = 1$}
When $\alpha = 1$, there are no terms in the summation, as the first $\int \cosh^q\psi \, {\rm d}\psi$ integral has $q = 0$. The solution is then 
\beq
f_1(x) = {x \over 2}\sqrt{1 - {2 \over x}} - \arccosh\sqrt{x\over2}
\eeq

\subsubsection{$\alpha = n + 1/2$, where $n \in \mathbb{Z}^+$}
When $\alpha = n + 1/2$, for positive integer $n$, the summation terminates at $k = n$, and we take ${\cal R}_\alpha = 0$. Our solution is then 
\beq
f_\alpha = {x^\alpha \over 2 \alpha} \sqrt{1 - {2\over x}} \left[ 1 - {1\over 2\alpha - 1} \sum_{k = 0}^n \beta_k \left({2\over x}\right)^{k+1}\right] . 
\eeq

\subsubsection{$\alpha = n+1$, where $n \in \mathbb{Z}^+$}
When $\alpha = n+1$, and is a positive integer greater than or equal to 2, the summation appears to diverge at $k = n$. However, this simply results from the use of the integral reduction formula for $\int \cosh^q\psi \, {\rm d}\psi$ when $q=0$, where it is not valid.  The summation must therefore be terminated at $n - 1$, with ${\cal R}_\alpha = 0$ and the final term replaced with $\arccosh\sqrt{x/ 2}$.  Explicitly 
\begin{multline}
f_\alpha = {x^\alpha \over 2 \alpha} \sqrt{1 - {2\over x}} \left[ 1 - {1\over 2\alpha - 1} \sum_{k = 0}^{n-2} \beta_k \left({2\over x}\right)^{k+1}\right] \\- {2^\alpha \over 2\alpha(2\alpha - 1)} \beta_{n-1} \arccosh\sqrt{x\over 2} . 
\end{multline}

\subsection{Limiting behaviour}
\subsubsection{$x \gg 2$}
When $x \gg 2$, the leading order behaviour of $f_\alpha$ is simple, as all terms in the summation are subdominant to the leading $x^\alpha$ term:
\beq
f_\alpha(x\gg2) = {x^\alpha \over 2\alpha} \sqrt{1 - {2\over x}} \left[ 1 + {\cal O}\left({2 \over x} \right) \right]
\eeq
\subsubsection{$x \rightarrow 2$}
Some care is required in understanding the $x \rightarrow 2$ behaviour of our general solution.  This involves analysing the next to leading order term in equation \ref{limit}:  
\begin{multline}
\lim_{x\rightarrow 2} \left( \sqrt{1-{2\over x}}  {}_2F_1\left(1, {3\over 2}-\alpha; 2-\alpha; {2\over x}\right)\right) = \\ \sqrt{1- {2\over x}} {\Gamma(2-\alpha) \Gamma(-1/2) \over \Gamma(1-\alpha)\Gamma(1/2) }+ C = 2 (\alpha - 1)\sqrt{1- {2\over x}} + C,
\end{multline}
where the constant ${C}$ will cancel with ${\cal R}_\alpha$ (by definition). Therefore the limiting behaviour of $f_\alpha$ is 
\beq
f_\alpha(x\rightarrow 2) \sim \sqrt{1 - {2\over x}} \left[1 - {2\over x} + \dots   \right] \sim \left(1 - {2\over x}\right)^{3/2}
\eeq

\section{The spin independence of the general relativistic Green's function}\label{spin_stuff}

\begin{figure}
\includegraphics[width=\linewidth]{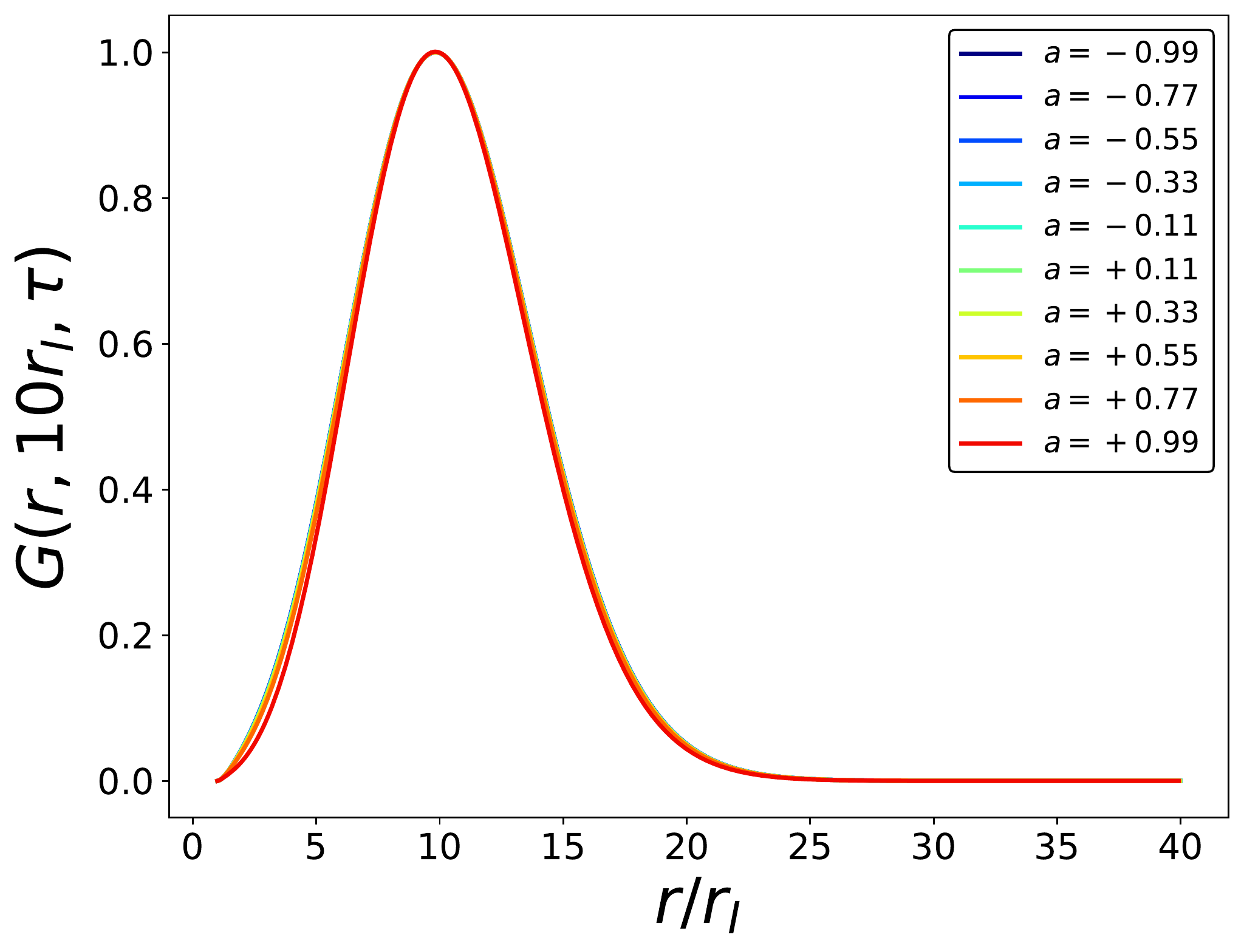}
\caption{Numerical solutions of the general relativistic disc equations, at a viscous time $t/t_{\rm visc} = 0.16$, for a number of different black hole spins displayed on the figure.  }
\label{a1}
\end{figure}

\begin{figure}
\includegraphics[width=\linewidth]{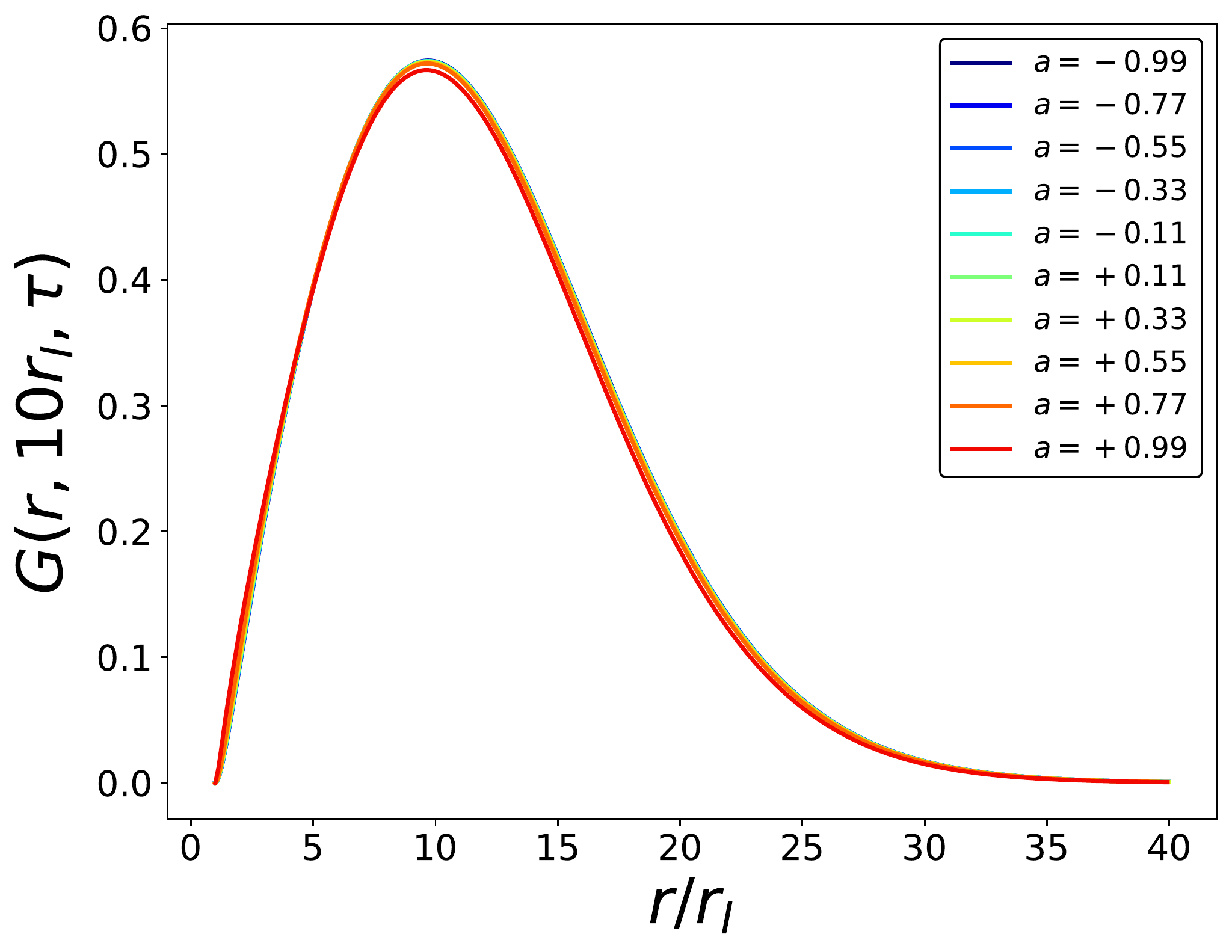}
\caption{Numerical solutions of the general relativistic disc equations, at a viscous time $t/t_{\rm visc} = 0.48$, for a number of different black hole spins displayed on the figure.  }
\label{a2}
\end{figure}

\begin{figure}
\includegraphics[width=\linewidth]{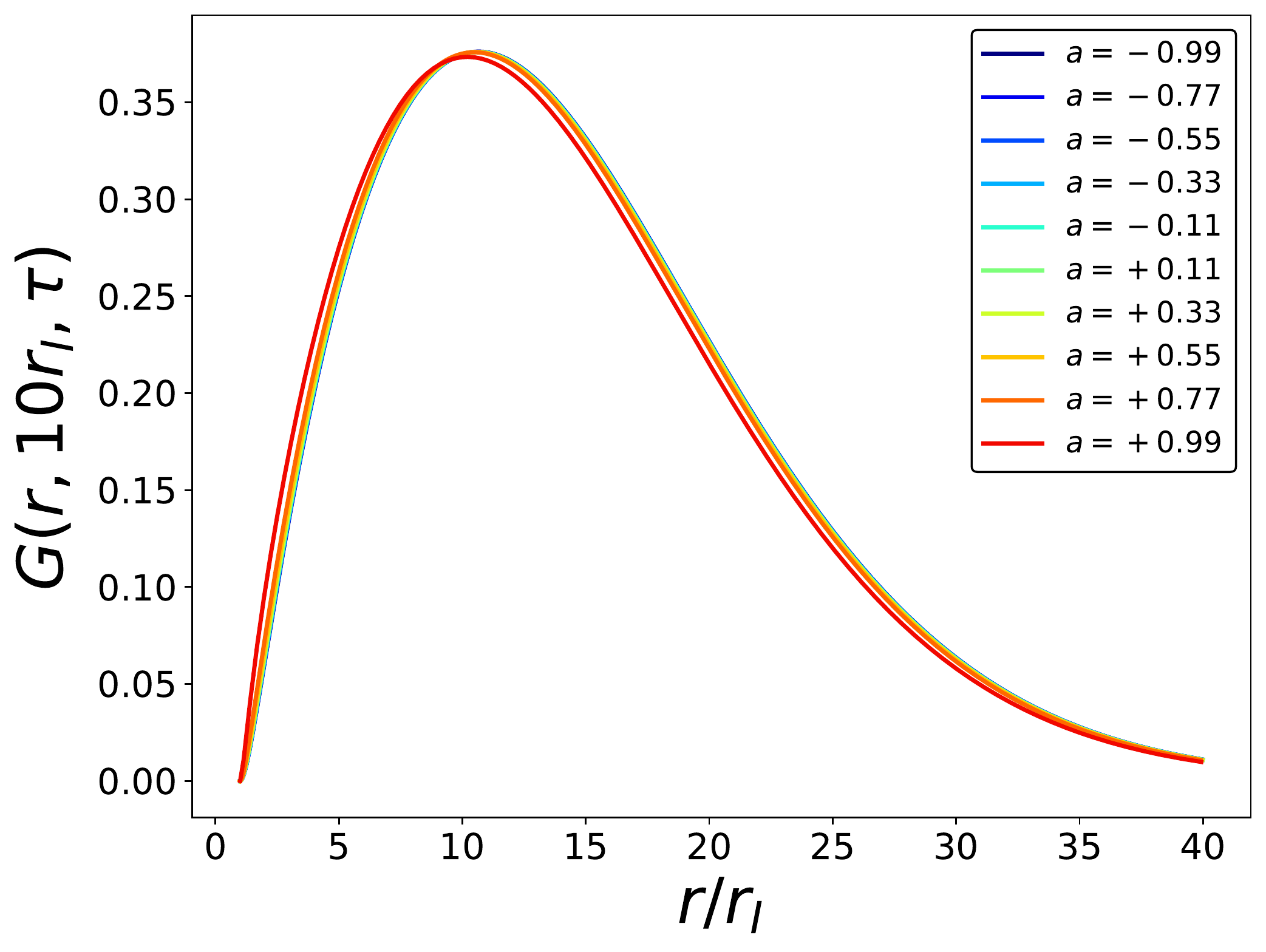}
\caption{Numerical solutions of the general relativistic disc equations, at a viscous time $t/t_{\rm visc} = 1.0$, for a number of different black hole spins displayed on the figure.  }
\label{a3}
\end{figure}

{The leading order Green's function solution presented in this paper (eq. \ref{green_y}) is independent of the black hole spin, once the radial coordinate is normalised by the ISCO radius. As a test of this analysis it is of interest to check whether the numerical solutions of the General Relativistic disc equations are themselves approximately independent of black hole spin, once the radius is suitably normalised. In Figures \ref{a1}, \ref{a2} and \ref{a3}, we compute the numerical Green's function solutions of equation (\ref{27q}), the full general relativistic disc equation,  for a variety of different values of the Kerr spin parameter, and plot them as a function of dimensionless radius $r/r_I$. Each numerical Green's function was initiated at a radius $r_0 = 10 r_I(a)$. }

{As can be clearly seen in Figures \ref{a1}, \ref{a2} and \ref{a3}, the gross behaviour of these numerical solutions is  well approximated as being independent of the black hole's spin, just as the analytical solution (eq. \ref{green_y}) suggests. }

\section{The derivative of the Green's function}\label{gradient}
We require the derivative  with respect to $x$ of the following Green's function 
\begin{multline}
G(x, x_0, \tau) = \sqrt{x^{-\alpha} f_\alpha(x) \exp\left({1 \over x} \right) \left(1 - {2\over x}\right)^{5/2 - 3/4\alpha} }\\ 
 {x^{1/4} \over \tau} \exp\left({-f_\alpha(x)^2 - f_\alpha(x_0)^2 \over  4\tau} \right) I_{1\over 4\alpha} \left({ f_\alpha(x) f_\alpha(x_0) \over  2\tau}\right),
\end{multline}
We will require the following Bessel function identity (Gradshteyn and Ryzhik {\it et al}. 2007)
\beq
{{\rm d} \over {\rm d}z} I_l(z) = I_{l-1}(z) - {l \over z} I_l(z).
\eeq
For notational ease I define 
\beq
F(x) = x^{{1/4-\alpha/2}} \exp\left({1 \over 2 x}\right) \left(1 - {2\over x}\right)^{5/4 - 3/8\alpha} ,
\eeq
such that
\begin{multline}
G(x, x_0, \tau) = {F(x) \sqrt{f_\alpha(x)} \over \tau} \exp\left({-f_\alpha(x)^2 - f_\alpha(x_0)^2 \over  4\tau} \right) \\ I_{1\over 4\alpha} \left({ f_\alpha(x) f_\alpha(x_0) \over  2\tau}\right) .
\end{multline}
The derivative we require is 
\begin{multline}
{\partial G \over \partial x} = {F(x) \sqrt{f_\alpha(x)} \over \tau} \exp\left({-f_\alpha(x)^2 - f_\alpha(x_0)^2 \over  4\tau} \right) \\ 
\Bigg[ \Bigg( {\partial \ln F \over \partial x} + {1 \over 2 f_\alpha(x)} {\partial f_\alpha \over \partial x} - {f_\alpha(x) \over 2 \tau} \Bigg)I_{1\over 4\alpha} \left({ f_\alpha(x) f_\alpha(x_0) \over  2\tau}\right) \\
+ { f_\alpha(x_0) \over 2\tau} {\partial f_\alpha \over \partial x} \Bigg[ I_{1 - 4\alpha\over 4\alpha} \left({ f_\alpha(x) f_\alpha(x_0) \over  2\tau}\right) \\ 
- {1\over 4\alpha} {2\tau \over f_\alpha(x) f_\alpha(x_0)} I_{1\over 4\alpha} \left({ f_\alpha(x) f_\alpha(x_0) \over  2\tau}\right)\Bigg)\Bigg]
\end{multline}
Using 
\beq
{\partial \ln F\over \partial x} = {1 - 2\alpha \over 4x} - {1\over 2 x^2} + {10\alpha - 3 \over 4 \alpha x^2 } {1 \over 1 - 2/x} 
\eeq
and noting that as $f_\alpha$ is defined via an integral its derivative is trivial 
\beq
{\partial f_\alpha \over \partial x} = {1\over 2} x^{\alpha - 1} \sqrt{1 - {2\over x}} ,
\eeq
leads to the result stated in the paper. 

\section{Numerical comparison to the Newtonian Green's functions}\label{obvious_stuff}
\begin{figure}
\includegraphics[width=\linewidth]{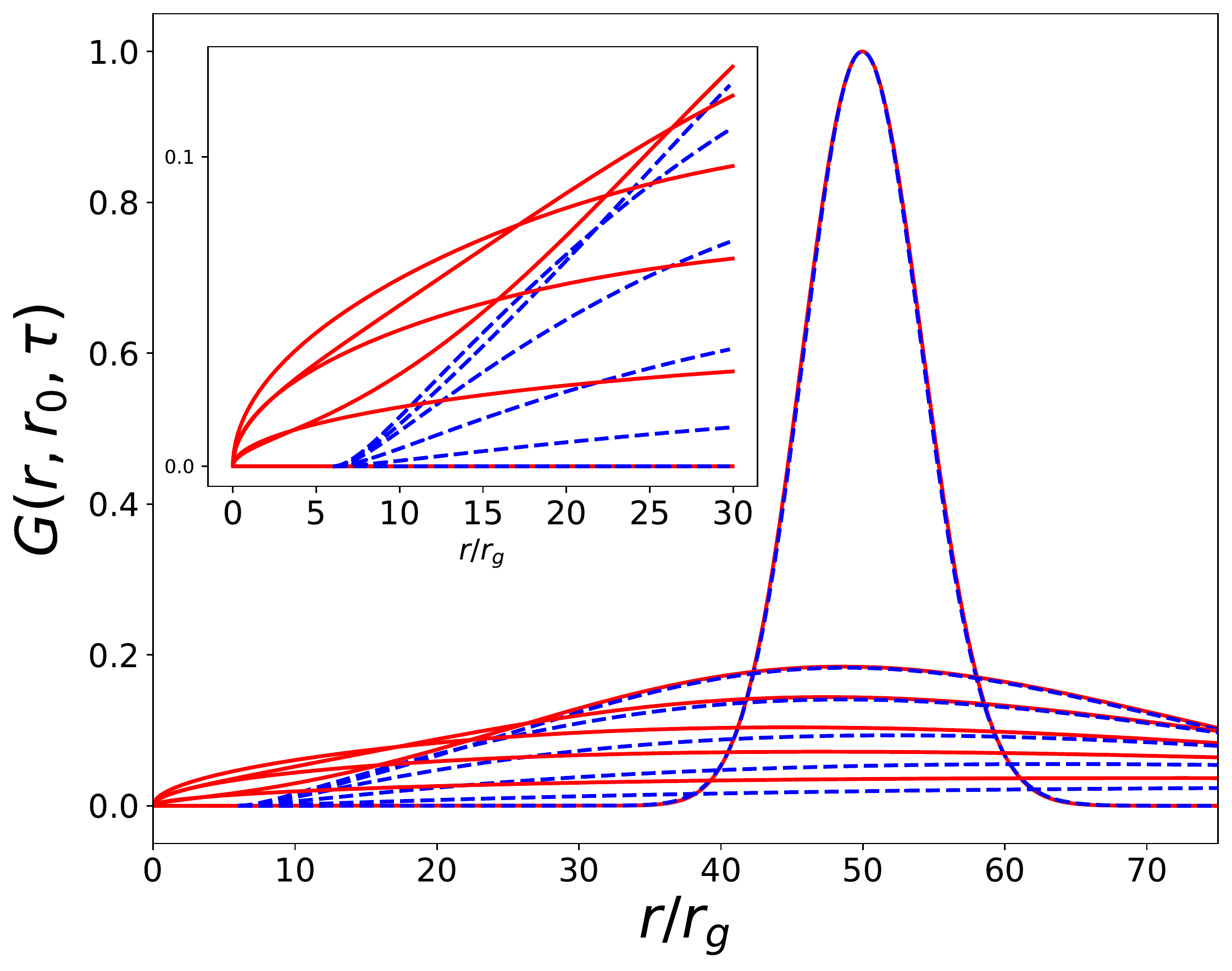}
\caption{Numerical solutions of the Schwarzschild disc equations, at viscous times $t/t_{\rm visc} = 0.008, 0.24, 0.4, 0.8, 1.6$ and $4$, versus the Newtonian (Lynden-Bell \& Pringle 1974) Green's functions. The initial radius of the Green's functions was chosen to be $r_0 = 50r_g$, the same as in Figures  \ref{max_spin_test},  \ref{zero_spin_test} and \ref{minus_max_spin_test}  in the main text.  The upper left panel shows a zoom in of the inner 30 gravitational radii. The Newtonian Green's functions are a poor approximation to the relativistic solutions  at small radii and early times, and at all radii at large times. }
\label{a4}
\end{figure}

{In this appendix we test whether the Green's function solutions derived in this paper provide a significant improvement when compared to the classical Newtonian (Lynden-Bell \& Pringle 1974) Green's functions. In Figure \ref{a4} we plot (blue dashed curves) the evolution of the numerical Schwarzschild disc equations with initial radius $r_0 = 50r_g$ (i.e., the solutions presented in Figure \ref{zero_spin_test}). Also plotted are the Newtonian Green's function solutions (red solid curves). The upper left panel of Fig. \ref{a4} shows a zoom in of the inner 30 gravitational radii. 

Unsurprisingly, at early times the Newtonian Green's functions provide a poor approximation to the true relativistic solutions in the inner $\lesssim 30$ gravitational radii, but are a reasonable description of the numerical solutions at large radii $r \gtrsim r_0$.  However, at later times $t \gtrsim t_{\rm visc}$ the Newtonian solutions deviate from the full numerical solutions {\it at all radii}. This is clearly demonstrated in the final three curves in Fig. \ref{a4}, where it can be seen that the Newtonian Green's function solutions overestimate the remaining surface density of the disc, a behaviour which persists out to very large radii $r \sim 100 r_g$. }

\label{lastpage}

\begin{thebibliography}{99}
\bibitem{}
Abramowitz M., Stegun I.~A., 1965, hmfw.book
\bibitem{}
Balbus S.~A., Hawley J.~F., 1991, ApJ, 376, 214. 
\bibitem{}
Balbus S.~A., 2017, MNRAS, 471, 4832
\bibitem{}
Balbus S.~A., Mummery A., 2018, MNRAS, 481, 3348.
\bibitem{}
Bender C.~M., Orszag S.~A., 1978
\bibitem{}
Eardley D.~M., Lightman A.~P., 1975, ApJ, 200, 187
\bibitem{}
Gradshteyn I.~S., Ryzhik I.~M., Jeffrey A., Zwillinger D., 2007
\bibitem{}
Klu{\'z}niak W., Lee W.~H., 2002, MNRAS, 335, L29.
\bibitem{}
Lynden-Bell D., Pringle J.~E., 1974, MNRAS, 168, 603
\bibitem{}
Lyubarskii Y.~E., 1997, MNRAS, 292, 679
\bibitem{}
Mummery A., Balbus S.~A., 2019, MNRAS, 489, 143
\bibitem{}
Mushtukov A.~A., Ingram A., van der Klis M., 2018, MNRAS, 474, 2259
\bibitem{}
Paczy{\'n}sky B., Wiita P.~J., 1980, A\&A, 88, 23
\bibitem{}
Shakura N.~I., Sunyaev R.~A., 1973, A\&A, 24, 337
\end{thebibliography}
\end{document}